\documentclass{jpp}

\usepackage{amsfonts}
\usepackage{amsmath}
\usepackage{times}
\usepackage{graphicx}
\usepackage{natbib}

\renewcommand{\vec}[1]{\mbox{\boldmath $#1$}}
\def \i{{\rm i}}

\def \d {{\rm d}}
\def \Om {{\it \Omega}}
\def \Omout {{\it \Omega_{\rm out}}}
\def \OmA {{\it \Omega_{\rm A}}}
\def \OmAf {{\it \Omega^\phi_{\rm A}}}
\def \OmAz {{\it \Omega^z_{\rm A}}}
\def \mur {r_{\rm in}}
\def \rin {r_{\rm in}}
\def \Re {\ensuremath{\rm{Re}}}
\def \Ha {\ensuremath{\rm{Ha}}}
\def \Hamin {\ensuremath{\rm{Ha_{min}}}}
\def \Pm {\ensuremath{\rm{Pm}}}
\def \Rm {\ensuremath{\rm{Rm}}}
\def \Mm {\ensuremath{\rm{Mm}}}
\def \S {\ensuremath{\rm{S}}}

\def \omdr{\omega_{\rm dr}}

\def\A{{Alfv\'en}}

\newcommand{\gsim}{\lower.7ex\hbox{$\;\stackrel{\textstyle>}{\sim}\;$}}
\newcommand{\lsim}{\lower.7ex\hbox{$\;\stackrel{\textstyle<}{\sim}\;$}}

\def\ara\&a{ Ann. Rev. Astronomy Astrophysics}

\shorttitle{Magnetic Taylor-Couette flows with super-rotation}
\title{Destabilization of super-rotating Taylor-Couette flows by current-free helical magnetic fields}
\author{G. R\"udiger\aff{1,2}
 \corresp{\email{gruediger@aip.de}}, 
 M. Schultz\aff{2}
 \and R. Hollerbach\aff{3} 
 }
\affiliation
{\aff{1}University of Potsdam, Institute of Physics and Astronomy, Karl-Liebknecht-Str.~24-25, 14476 Potsdam, Germany
\aff{2}Leibniz-Institut f\"ur Astrophysik Potsdam, An der Sternwarte 16, D-14482 Potsdam, Germany
\aff{3}Department of Applied Mathematics, University of Leeds,
Leeds, LS2 9JT, United Kingdom
}

\date{\today}
\begin{document}

\maketitle

\begin{abstract}
In an earlier paper we showed that the combination of azimuthal magnetic fields and super-rotation in Taylor-Couette flows of conducting fluids can be unstable against non-axisymmetric perturbations if the magnetic Prandtl number of the fluid is $\Pm\neq 1$. Here we demonstrate that the addition of a weak axial field component may allow axisymmetric perturbation patterns for $\Pm$ of order unity depending on the boundary conditions. The axisymmetric modes only occur for magnetic Mach numbers (of the azimuthal field) of order unity, while higher values are necessary for non-axisymmetric modes. The typical growth time of the instability and the characteristic time scale of the axial migration of the axisymmetric mode are long compared with the rotation period, but short compared with the magnetic diffusion time. The modes travel in the positive or negative $z$-direction along the rotation axis depending on the sign of $B_\phi B_z$. We also demonstrate that the azimuthal components of flow and field perturbations travel in phase if $|B_\phi|\gg |B_z|$, independent of the form of the rotation law. Within a short-wave approximation for thin gaps it is also shown (in an Appendix) that for {\em ideal} fluids the considered helical magnetorotational instability (HMRI) only exists for rotation laws with negative shear.
\end{abstract}

\maketitle

\section{Introduction}
Cylindrical Taylor-Couette containers filled with a conducting fluid and subject to externally applied large-scale magnetic fields can be used as a `virtual' laboratory to study magnetic instabilities. The simplest geometry of the external magnetic field is a homogeneous axial field, reproducing the standard magnetorotational instability (MRI) if the outer cylinder rotates at a slower frequency than the inner one. \cite{V59} showed the instability of this constellation for ideal flows. By including diffusive effects, \cite{JG01} and \cite{RZ01} started to probe the Taylor-Couette flow as the appropriate object to study the (present-day) variety of  magnetic instabilities by means of theory, experiments and numerical simulations. The  references describing the detailed  history of hydromagnetic Taylor-Couette research and the corresponding laboratory experiments are given in the recent  review  \citep{RGH18}.

Axisymmetric as well as  non-axisymmetric perturbation patterns are both unstable, with axisymmetric modes excited first, that is, at slower rotation rates. Due to diffusion, this instability requires a minimum magnetic field for excitation, with a critical Lundquist number $\S\simeq 1$ (see below for the exact definitions of parameters such as $\S$).

Once the axisymmetric mode (``channel flow'') is excited, any further increase of the Reynolds number $\Re$ does not restabilize the flow. This, however, is not true for the non-axisymmetric modes, which can always be restabilized by faster rotation. The stability maps for non-axisymmetric modes of a fluid rotating beyond the Rayleigh limit (e.g.\ quasi-Keplerian rotation) show for any given Lundquist number of the axial magnetic field a lower  critical Reynolds number for the MRI onset and a maximal  one  where the diffusion stops the instability again \citep{GR12}.  The minimum rotation rates of the lines of neutral stability scale with $\Pm\Re\simeq$~const for small $\Pm$, and with $\sqrt{\Pm}\ \Re \simeq$~const for large $\Pm$, where the magnetic Prandtl number
\begin{eqnarray}
 {\Pm} = \frac{\nu}{\eta}
 \label{pm0}
\end{eqnarray}
is the ratio of kinematic viscosity $\nu$ and magnetic diffusivity $\eta$. Hence, the  lowest critical  rotation rates $\Om$ scale as $\Om\propto \eta$ for $\Pm\ll1$ and as $\Om\propto \sqrt{\nu\eta}$ for $\Pm\gg 1$.  They are obviously minimal for $\Pm$ of order unity. This well-known standard type of MRI does {\em not} exist for rotation profiles with positive shear (super-rotation).

Another type of MRI appears if the applied magnetic field is azimuthal and curl-free in the gap between the cylinders. This configuration exhibits only non-axisymmetric instability modes, but independent of the sign of the shear of the rotation. There is again a minimum Reynolds number for excitation, but unlike the standard MRI, the azimuthal magnetorotational instability (AMRI) is suppressed again if the rotation is too rapid. The Hartmann number $\Ha$ exhibits the same behaviour, with a minimum value required, but the AMRI is also suppressed again if the applied field is too strong. The lines of neutral stability of these modes thus form typical oblique cones in the ($\Ha/\Re$) plane, where the slopes $\d \Re/\d \Ha$ of the two branches are positive, and the Hartmann number $\Hamin$ at the point where $\d \Re/\d \Ha=\infty$ defining the overall weakest magnetic field amplitude for instability.

A very special situation holds for AMRI flows with super-rotation, when the outer cylinder rotates with a higher frequency than the inner one. For small magnetic Prandtl numbers the lines of neutral stability coincide in the ($\Ha/\Re$) plane, whereas for large $\Pm$ they coincide in the ($\Ha/\Rm$) plane, where $\Rm=\Pm\Re$ is the magnetic Reynolds number. One might not expect problems in the limit $\Pm\to 1$ but they do exist. Approaching $\Pm=1$, the critical values for both $\Ha$ and $\Re$ go to infinity, for both $\Pm<1$ and $\Pm>1$. The magnetized flow for $\Pm=1$ is stable, but is unstable for $\Pm\neq 1$; that is, this is a so-called double diffusive instability. We have numerically demonstrated this behaviour of the critical values for a container with an almost stationary inner cylinder \citep{RSG18}. The possible existence of solutions for $\Pm=1$ is of particular relevance if turbulent fluids are considered, as the effective magnetic Prandtl number in turbulent media basically approaches unity. 

The present paper addresses the problem of how the characteristics of this instability for azimuthal field $B_\phi$ and super-rotation are modified if the azimuthal field is complemented by a small axial component $B_z$. The resulting field then possesses a helical structure, as we  considered earlier but with sub-rotation \citep{HR05,SG06}. Even for { ideal fluids}, with vanishing diffusivities, the latter constellation under the presence of non-uniform rotation with negative shear proves to be unstable against axisymmetric perturbations. By use of a short-wave approximation -- which should be applicable for flows without narrow gaps -- we shall  show in the below Appendix   that the helical field becomes overstable under the presence of differential rotation with negative shear. The dimensional growth rate of the resulting axial wave basically scales with \A~frequency of the axial field. The influence of the azimuthal field on the growth rate proves to be negligible. For { real fluids} with finite  diffusivities the stability maps in the ($\Ha/\Re$) plane are  very similar to those of the standard MRI. For all not too small Hartmann numbers (formed with the axial field) there exists a critical Reynolds number above which the system is unstable against axisymmetric perturbations. At a certain Hartmann number the critical Reynolds number always  possesses a minimum, which for conducting cylinders is much lower than for insulating ones \citep{RGH18}.

But the short-wave approximation does {\em not} provide positive growth rates for ideal flows rotating with positive shear. Helical magnetic fields are thus stable  in super-rotating ideal  fluids. We have to underline, however,  that our (numerical) proof only bases on a short-wave approximation. One finds the similar situation for the nonaxisymmetric AMRI: it exists for ideal fluids only for sub-rotation rather than for super-rotation. In the latter case a possible instability must be diffusion-driven. Indeed, the simulations reveal current-free azimuthal magnetic fields as unstable under the influence of super-rotation but only for finite diffusivities under the condition $\nu\neq \eta$. Figure 5 of \cite{RGH18} demonstrates for perfectly conducting cylinders how the critical magnetic fields and the rotation rates go to infinity if $\Pm\to 1$. This is a typical behaviour for double-diffusive instabilities \citep{A78,K13,K17}.

We shall here demonstrate that for helical background fields the instabilities are still qualitatively of a double-diffusive character, in the sense that they operate most efficiently for either $\Pm\ll1$ or $\Pm\gg1$. In some aspects though they are also different from a classical double-diffusive type of instability, which would require the `singularity' to occur at precisely $\Pm=1$ \citep{K17}. Instead, we find here that the behavior depends on the imposed boundary conditions: For perfectly conducting boundaries there is indeed a range of $\Pm$ values for which no instability at all occurs, and this range includes $\Pm=1$. In contrast, for insulating boundaries there is no such gap in $\Pm$; around $\Pm\approx0.28$ there is instead a regime where both the `small-$\Pm$' and `large-$\Pm$' branches have large but still finite critical Hartmann and Reynolds numbers. In this scenario instabilities therefore do exist for $\Pm=1$, and are already part of the `large-$\Pm$' branch. This unexpected result that the behavior in the range $\Pm=O(1)$ depends so crucially on the choice of boundary conditions underlines  that the  simplification  of $\Pm=1$ often used in theories and simulations has its own risks. By use of a short-wave approximation \cite{MS19} demonstrate that super-rotating helical magnetic fields should never be unstable for $\Pm=1$. See also \cite{K13} for discussion of how boundary conditions can influence a range of stability problems generally, and how such effects can be analysed.

A strong influence  of the boundary conditions on critical Hartmann and Reynolds numbers also exists for  AMRI flows with super-rotation if $\Pm\ll 1$. In the inductionless limit $\Pm=0$   even the  minimal shear values defined by \cite{LG06,KS12}  differ for differing boundary conditions \citep{RSG18}. Note also that here we consider only radial boundary conditions, with the system assumed to be periodic and thus infinite in the axial direction. Any actual experiment of course would necessarily be finite in height, and whether the resulting axial boundaries are insulating or conducting can also play an important role \citep{CC18,CE19}.

The basic parameter in this study is the ratio of the azimuthal to the axial field component,
\begin{eqnarray}
\beta=\frac{B_\phi(R_{\rm in})}{B_z},
\label{beta}
\end{eqnarray}
where $R_{\rm in}$ is the radius of the inner cylinder. We are interested in the limit where $\beta$ is large, and can be positive or negative. For comparison, in the solar convection zone the equivalent $\beta$ is of order $10^3$, and is negative in the northern hemisphere, and positive in the southern.

It has been suggested that the migration toward the equator of the latitude of maximal solar activity over 11 years (the solar cycle) might be understood as a drifting axisymmetric mode of a magnetic instability driven by the super-rotation which exists beneath the equator at the bottom of the convection zone \citep{MS19}. The observations provide   another challenge to discuss the travelling magnetic instability patterns. During the activity cycle an azimuthal magnetic field band migrates from mid-latitudes toward the equator. At the equatorial side of this magnetic band there is a region of faster-than-average rotation, while at its pole-ward side there is a region of slower-than-average rotation \citep{KHH16}. The waves of azimuthal flow and azimuthal field, therefore,  may be  travelling out of phase in the Sun. We shall thus discuss for which rotation profiles and for which helicity-type (right-hand or left-hand) of the background field the waves of field and flow perturbations propagate in phase or out of phase. We  note  that during the 11-year solar cycle the Sun rotates 160 times, hence the ratio of the drift frequency of a hypothetical  magnetic wave  to the rotation frequency would be  $3\cdot 10^{-3}$.

The paper is structured as follows. The basic differential equations and boundary conditions are formulated in the following Section. In Section 3 the lines of marginal stability of the linearized system for various container sizes and for fixed $\Pm=1$ and $|\beta|=25$ are discussed. One finds axisymmetric and non-axisymmetric modes to be unstable, where the latter requires stronger fields and faster rotation for excitation. For the axisymmetric mode the inclination angle $\beta$ and the magnetic Prandtl number $\Pm$ are varied in Sections 4 and 5, where the axial drift relations in dependence on the sign of $\beta$ are also demonstrated. In the final sections the phase relations of the azimuthal components of flow and field for super-rotation and sub-rotation will be discussed. The results are reviewed in the last Section, where their possible connection to the cyclic activity of the Sun will also be discussed.

\section{The Equations}
The equations of the problem are
\begin{eqnarray}
 \frac{\partial \vec{U}}{\partial t} + (\vec{U}\cdot \nabla)\vec{U} &=& -\frac{1}{\rho} \nabla P + \nu \Delta \vec{U} 
 + \frac{1}{\mu_0\rho}{\textrm{curl}}\vec{B} \times \vec{B},\nonumber\\
 \frac{\partial \vec{B}}{\partial t}&=& {\textrm{curl}} (\vec{U} \times \vec{B}) + \eta \Delta\vec{B} 
 \label{mhd2}
\end{eqnarray}
with $ {\textrm{div}}\ \vec{U} = {\textrm{div}}\ \vec{B} = 0$ for an incompressible fluid.
$\vec{U}$ is the velocity, $\vec{B}$ the magnetic field, $P$ the pressure and $\rho$ the density. The basic state in the cylindrical system with coordinates $(R,\phi,z)$ is \mbox{$ U_R=U_z=B_R=0$} for the poloidal components and \mbox{$\Om = a_\Om + {b_\Om}/{R^2}$}  with 
\begin{eqnarray}
 a_\Om=\frac{\mu-\mur^2}{1-\mur^2}\Om_{\rm in}, \quad
 b_\Om= \frac{1-\mu}{1-\mur^2}\Om_{\rm in} R_{\rm in}^2,
 \label{ab} 
\end{eqnarray}
where $\mur={R_{\rm in}}/{R_{\rm out}}$ is the ratio of the two cylinders' radii, and $\Om_{\rm in}$ and $\Om_{\rm out}$ are their angular velocities. If we define the ratio $\mu= \Om_{\rm out}/\Om_{\rm in}$, then super-rotation is represented by \mbox{$\mu>1$}.

The current-free azimuthal field is given by
$
 B_\phi=R_{\rm in}B/R.
$
Together with a uniform axial component $B_z$, the externally imposed field is therefore  $\vec{B}=(0, R_{\rm in} B/R, B_z)$.

In addition to the magnetic Prandtl number $\Pm$, which is a material property of the fluid, the other dimensionless parameters of the system are the Hartmann number $\Ha$ and the Reynolds number $\Re$, 
\begin{eqnarray}
 {\Ha} =\frac{B_{z} R_0}{\sqrt{\mu_0\rho\nu\eta}}, \quad\quad\quad
 {\Re} =\frac{\Om_{\rm out} R_0^2}{\nu},
\label{pm}
\end{eqnarray}
which measure the strength of the imposed axial field and the outer cylinder's rotation rate, respectively. Alternative measures are the Lundquist number $\S= \sqrt{\Pm}\ \Ha$ and the magnetic Reynolds number $\Rm=\Pm\ \Re$. Different choices of $\Ha$ versus $\S$, and $\Re$ versus $\Rm$, are appropriate in different limiting parameter regimes. The parameter $R_0=\sqrt{(R_{\rm out}-R_{\rm in})R_{\rm in}}$ is a suitably scaled measure of length.

Recalling the ratio  (\ref{beta}), it is useful to also define an azimuthal Hartmann number
 $\Ha_\phi=\beta \Ha$, which measures the strength of the azimuthal field $B_\phi$ rather than the axial field $B_z$. These quantities may be combined to yield the magnetic Mach number of the {\em azimuthal} field, 
\begin{eqnarray}
\Mm=\sqrt{\Pm}\frac{\Re}{\beta\Ha},
\label{Mm}
\end{eqnarray}
measuring whether the rotation energy dominates the magnetic energy or not. The magnetic Mach number of cosmical objects almost always exceeds unity. Adopting solar values, ($U_\phi\simeq 2$~km/s, $B_\phi\simeq 1$~kG), one finds ${\Mm}\simeq 500 \sqrt{\rin(1-\rin)}$, which already exceeds unity for
the very small gap width of $3$~km. For the solar tachocline with its thickness of 50,000 km, the magnetic Mach number is $\Mm\simeq 150$, or $\Mm\simeq 15$ for the stronger azimuthal field $B_\phi\simeq 10$~kG.

The equations are linearized, and instability modes of the form $f=f(R){\rm{exp}}\bigl({\rm{i}}(kz+m\phi+\omega t)\bigr)$ are sought. The result is a linear, one-dimensional eigenvalue problem, with only the radial structures $f(R)$ still to be solved for, and with $\omega$ being the eigenvalue. This eigenvalue system is solved by finite-differencing in $R$, as in \cite{SR02}, or alternatively by Chebyshev expansions, as in \cite{HR05}. For a given Hartmann number, solutions are optimised with respect to the Reynolds number by varying the axial wave number $k$. The azimuthal wave number $m$ is either $0$ for axisymmetric modes, or $\pm1$ for non-axisymmetric modes. Higher non-axisymmetric modes can also be excited, but typically at higher Hartmann and/or Reynolds numbers than $m=\pm1$, so we focus on $m=0$ and $\pm1$ here. There are also various symmetries that apply to positive versus negative $m$. For purely azimuthal fields ($B_z=0)$, $m\to-m$ are directly equivalent, whereas for general helical fields $m\to-m$ are equivalent if additionally one takes either of $k\to-k$ or $\beta\to-\beta$. One can therefore restrict attention to either positive $m$ or positive $\beta$, for example, as long as the other one is allowed to take on both signs.

The associated boundary conditions are no-slip for the velocity perturbations, ${\vec u}=0$. For the boundary conditions on ${\vec b}$ one can take the cylinders to be either perfectly conducting or insulating. Conducting boundary conditions are ${\rm d} b_\phi/{\rm d}R + b_\phi/R=b_R=0$ at both $R_{\rm in}$ and $R_{\rm out}$. Insulating boundary conditions are  more complicated, and different at 
$R_{\rm in}$ and $R_{\rm out}$, i.e.\ for $R=R_{\rm in}$
\begin{equation}
b_R+ \frac{{\rm i} b_z}{I_m(kR)} \left(\frac{m}{kR} I_m(kR)+I_{m+1}(kR)\right)=0,
\label{72.7}
\end{equation}
and for $R=R_{\rm out}$ 
\begin{equation}
b_R+ \frac{{\rm i} b_z}{K_m(kR)} \left(\frac{m}{kR} K_m(kR)-K_{m+1}(kR)\right)=0,
\label{72.8}
\end{equation}
where $I_m$ and $K_m$ are the modified Bessel functions. (Note that these satisfy $I_{-m}=I_{m}$ and $K_{-m}=K_{m}$.) Additionally, the toroidal field at both boundaries must satisfy $k R\, b_\phi =m\, b_z$.  Note also that in all cases the total number of boundary conditions  correctly matches the number of equations in the eigenvalue problem. See also \cite{RSS18} for a detailed derivation of these radial boundary conditions, including the option of finitely conducting boundaries. 

The linear code works with length scales normalized with $R_0$, and frequencies normalized with $\Om_{\rm out}$. Positive values of the drift frequencies denote negative axial phase velocities, so that the instability pattern migrates in the negative $z$-direction, anti-parallel to the rotation axis. For negative drift frequencies it is vice versa. The drift frequency can also be normalized with the magnetic diffusion frequency 
\begin{equation}
\omega_{\rm diff}=\frac{\Om_{\rm out}}{\Rm}.
\label{omdiff}
\end{equation} 
Note finally that in any linear eigenvalue problem the overall solution amplitude is undetermined, so that only ratios of variables have clearly defined physical meanings.

We mainly deal with a flow with almost stationary inner cylinder, in narrow-gap configurations. We have earlier shown that a toroidal magnetic field which is current-free between the cylinders with the outer cylinder rotating faster than the inner cylinder may become unstable against non-axisymmetric perturbations with $m=\pm 1$ \citep{RGH18}. It is a double-diffusive instability which requires $\Pm\neq 1$ for its existence. It also exists in the inductionless approximation, $\Pm\to 0$, which automatically means that the relevant parameters for small magnetic Prandtl number are $\Re$ and $\Ha$. This is relevant for possible experiments with liquid metals with their very small $\Pm$ as for $\Pm\to 0$ the Reynolds number does not grow to infinity as is the case for instabilities where the relevant parameter is $\Rm$ rather than $\Re$. For $\Pm\gg1$ the rotational parameter scales with $\Rm$.

If an axial component is added to the imposed field, the first mode to go unstable becomes the axisymmetric $m=0$ mode. This is even true if the axial field is much smaller than the azimuthal field, i.e.\ for $\beta\gg 1$. For much smaller wave numbers the existence of another mode (`type 2') is reported which does not exist for $\Pm=1$  \citep{MS19}. It is a double-diffusive instability which lives from the difference of viscosity and resistivity, and which is not a solution of the MHD equations in the inductionless limit. 

\section{Axisymmetric and non-axisymmetric solutions for $\bf Pm=1$}
\begin{figure}
 \includegraphics[width=0.51\textwidth]{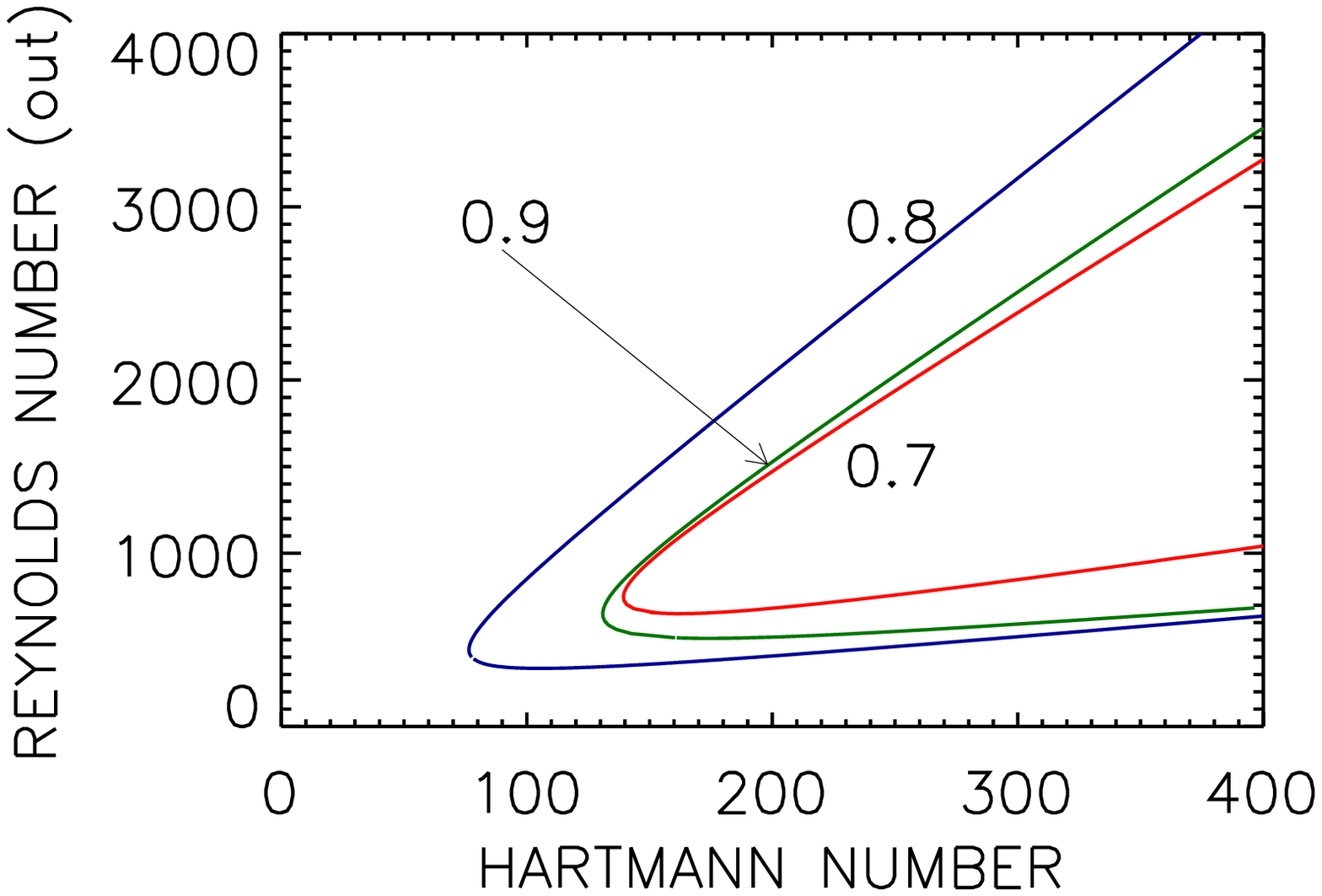}
 \includegraphics[width=0.51\textwidth]{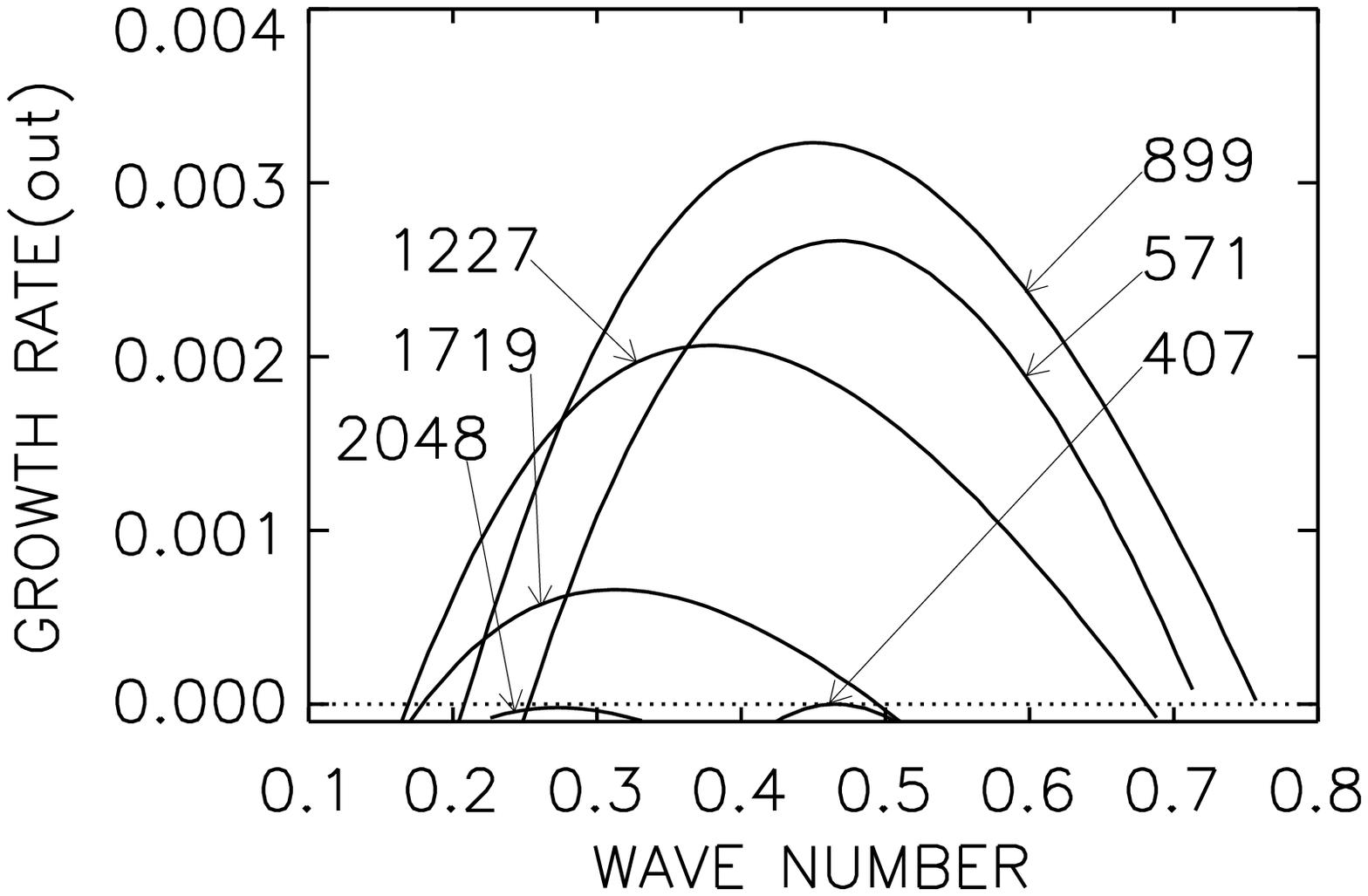}
 \caption{Left panel: Stability maps for the axisymmetric mode of the super-rotating flow.
 Green line: $\rin=0.9$; blue line: $\rin=0.8$; red line: $\rin=0.7$. Right panel: Growth rates normalized with the rotation rate of the outer cylinder for $\rin=0.8$ and  $\Ha=200$. The curves are marked with their Reynolds numbers. 
$m=0$, $\mu=128$, $\Pm=1$, $\beta=25$. Insulating cylinders.} 
 \label{figm0a} 
\end{figure}

In the next two  sections we shall consider the stability  of flows of magnetic Prandtl number unity between insulating wall, thus excluding all types of double-diffusive instabilities. We shall demonstrate that even in this case axisymmetric as well as non-axisymmetric perturbation modes are unstable for rotation laws with positive shear. We are mainly interested in magnetic background fields where the azimuthal component dominates; the general choice here is $\beta=25$. The gap width between the two cylinders is a free parameter, and we are interested in narrow gaps. The parameters which allow instability are the Reynolds number and the Hartmann number. They define an unstable domain which is limited by a lower and an upper Reynolds number, and similarly a lower and an upper Hartmann number. That is, the system is stable both for too slow and too fast rotation, and similarly for too weak and too strong fields, as seen in Fig.~\ref{figm0a}. The absolute minimal Hartmann number for marginal instability is called $\Hamin$.
\begin{figure}
\includegraphics[width=0.51\textwidth]{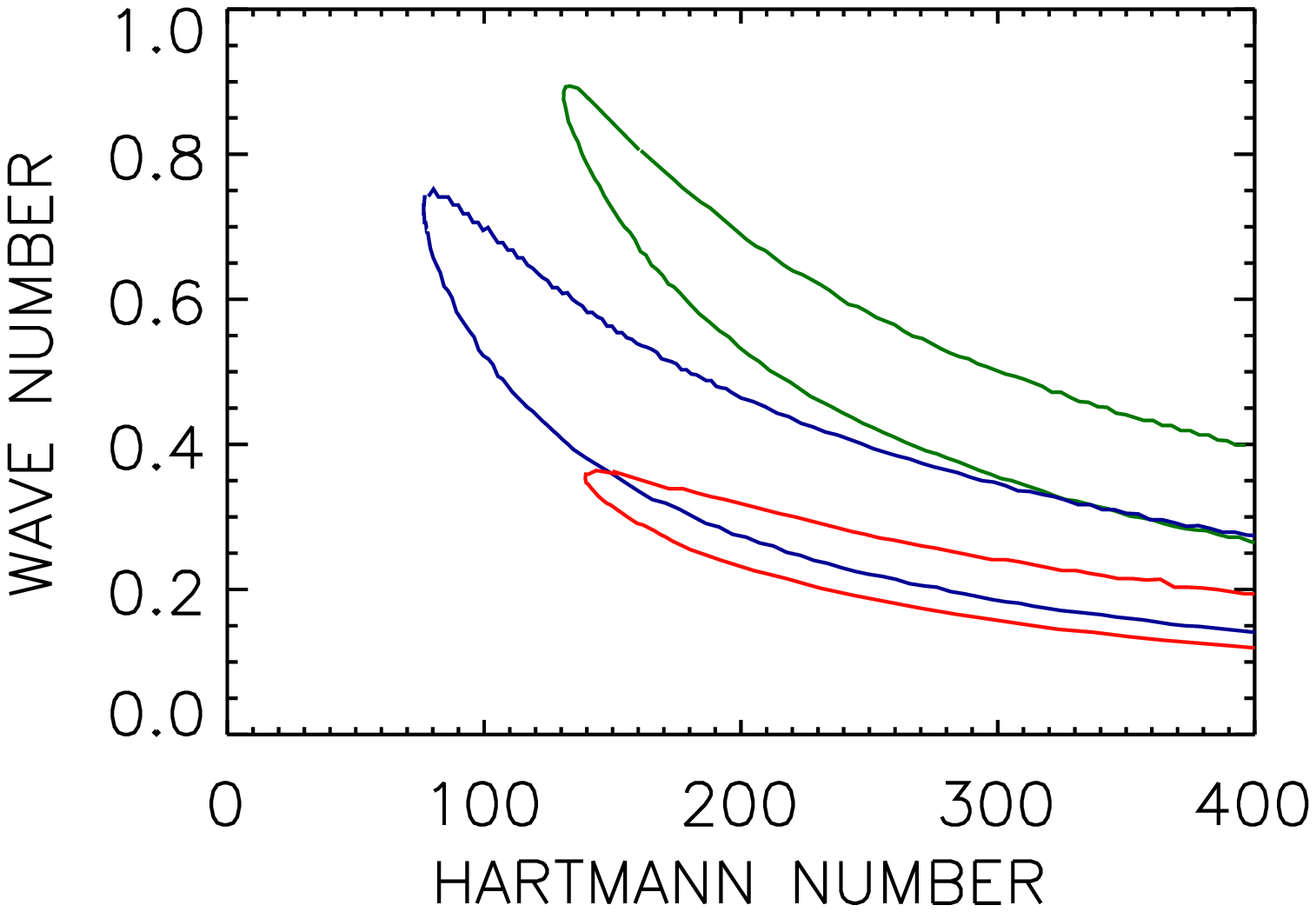}
 \includegraphics[width=0.51\textwidth]{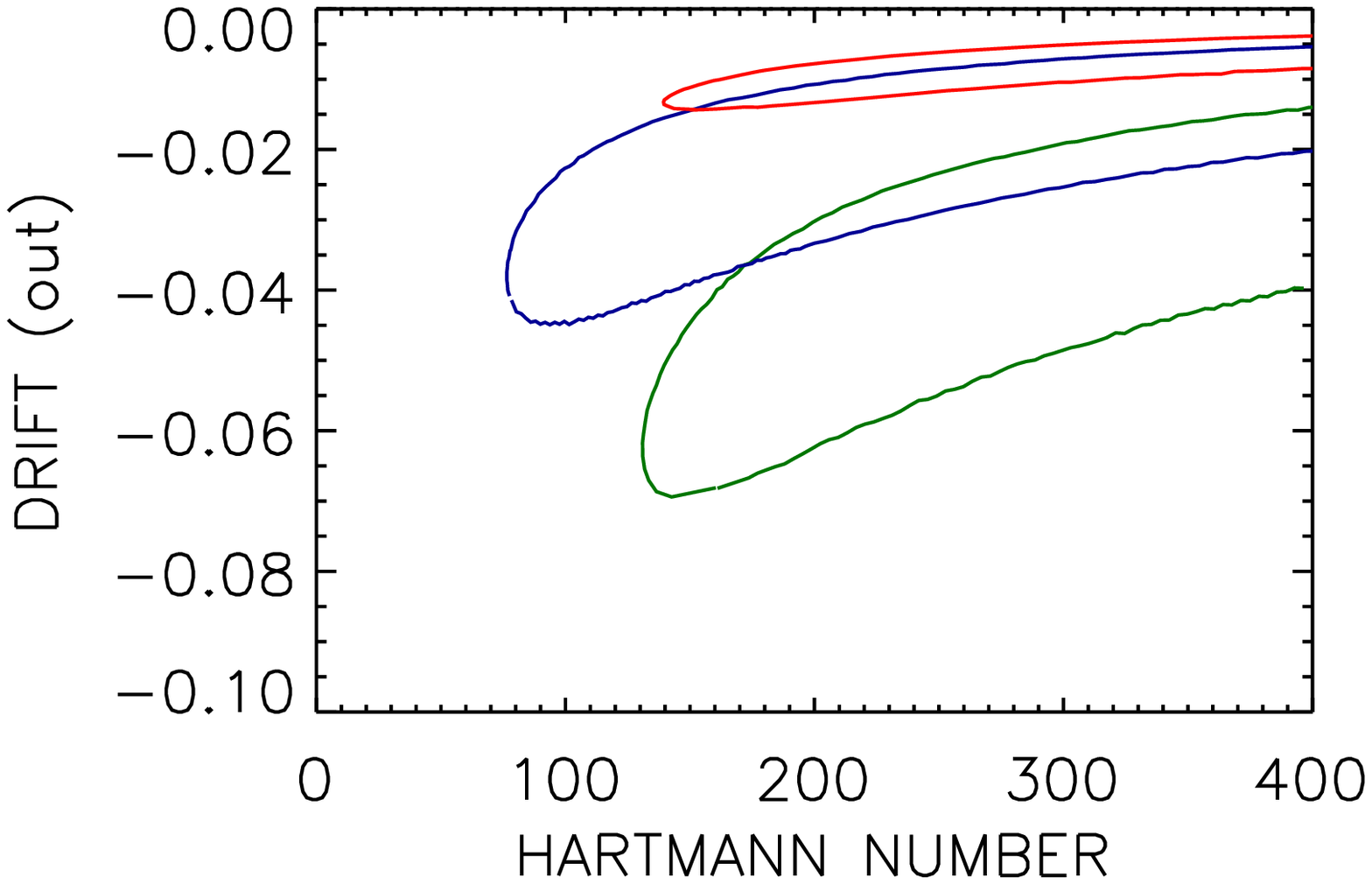}
 \caption{Similar to Fig.~\ref{figm0a} (left) but for normalized wave numbers  $k R_0$ (left panel) and drift frequencies (\ref{omdr}) (right panel).} 
 \label{figm0b} 
\end{figure}

For cylinders with $\rin=0.7$, $0.8$ and $0.9$, the left panel of Fig.~\ref{figm0a}  presents the lines of marginal stability of the axisymmetric perturbation modes in the $(\Ha/\Re)$ plane. The influence of the gap widths on the neutral stability lines is weak. One finds a minimal Hartmann number of order 100, with a weak dependence on the gap width. The instability only exists for magnetic Mach numbers (\ref{Mm}) -- of the azimuthal field --  beween 0.06 and 0.4. These values, defined with the azimuthal field amplitude, are strikingly small. Relative to the \A~frequency of the magnetic field the rotation rate must be rather {\em low} to destabilize the flow. Figure \ref{figm0a} also demonstrates that the instability only occurs in a  small part of the ($\Ha/\Re$) plane. The opening of the instability cone depends on the precise value of $\beta$. For $\beta\to 0$ and $\beta\to \infty$ the axisymmetric instability will disappear, so that one must expect that the opening of the cone becomes smaller and smaller for both decreasing and increasing $\beta$ (see below). For larger and larger Hartmann numbers the lines of marginal instability in Fig. \ref{figm0a} remain straight lines of constant slope, as we probed for the blue line up to $\Ha=10^4$. We did not find any indication of (island) instability domains, limited in their Hartmann numbers.

The right panel of Fig. \ref{figm0a} gives the dependence of the growth rate in units of the angular velocity of the outer cylinder for a fixed geometry ($\rin=0.8$) and a fixed Hartmann number ($\Ha=200$) as a function of the wave number $k R_0$ and Reynolds number $\Re$. One finds zero growth rates for the lower and the upper critical Reynolds numbers. Somewhere between these limits the growth rate becomes maximum at a wave number ($k R_0 \simeq 0.5$) close to that wave number where the instability sets on. The wave numbers and the maximum growth rates have very small values; the instability is thus slow also in comparison with the typical groth rates known for HMRI (see the below Fig. \ref{ideal2}) and/or AMRI with super-rotation \citep{RSG18}. Small wave numbers represent cells elongated in the axial direction.

The axial wave numbers (normalized with $R_0$) and the drift frequencies (normalized with the rotation frequency $\Omout$) along the neutral lines are given by  Fig.~\ref{figm0b}. By definition the cell size $\delta z$ along the rotation axis normalized with the gap width $D$ is $\delta z/D=\pi \sqrt{R_{\rm in}/D} /k R_0=  2\pi/kR_0$, the latter relation for $\rin=0.8$. Hence, a magnetic pattern which is nearly spherical in the gap between the cylinders should have a normalized wave number $k R_0 =2\pi$. The wave number values in the right panel of Fig. \ref{figm0a} and the left panel of Fig.~\ref{figm0b} are much smaller, so that the cells are instead rather long in the vertical direction $z$. For large Hartmann numbers the wave numbers decrease. The cells, therefore, become increasingly elongated for stronger axial fields, in agreement with the magnetic Proudman theorem.

The characteristic values of the drift 
\begin{eqnarray}
\omdr=\frac{\omega^{\rm R}}{\Omout},
\label{omdr}
\end{eqnarray}
where $\omega^{\rm R}$ is the real part of $\omega$, are negative and of order 0.05, which is {\em faster} than the diffusion drift $\omega_{\rm diff}\lsim 0.004$ of the magnetic pattern by one order of magnitude. More details of the drift phenomenon are  presented  in Section \ref{Axial}.

Non-axisymmetric modes are also unstable. The boundary conditions (\ref{72.7}) and (\ref{72.8}) for insulating cylinders also allow calculations for non-zero azimuthal wave numbers $\pm m$. One expects higher values for the excitation of non-axisymmetric modes if axisymmetric modes exist. The two solutions for $m=1$ and $m=-1$ form spirals of opposite chirality. The question is whether the two modes due to a background field with a fixed value of $\beta$ have different excitation  conditions  or not.

\begin{figure}
\includegraphics[width=0.51\textwidth]{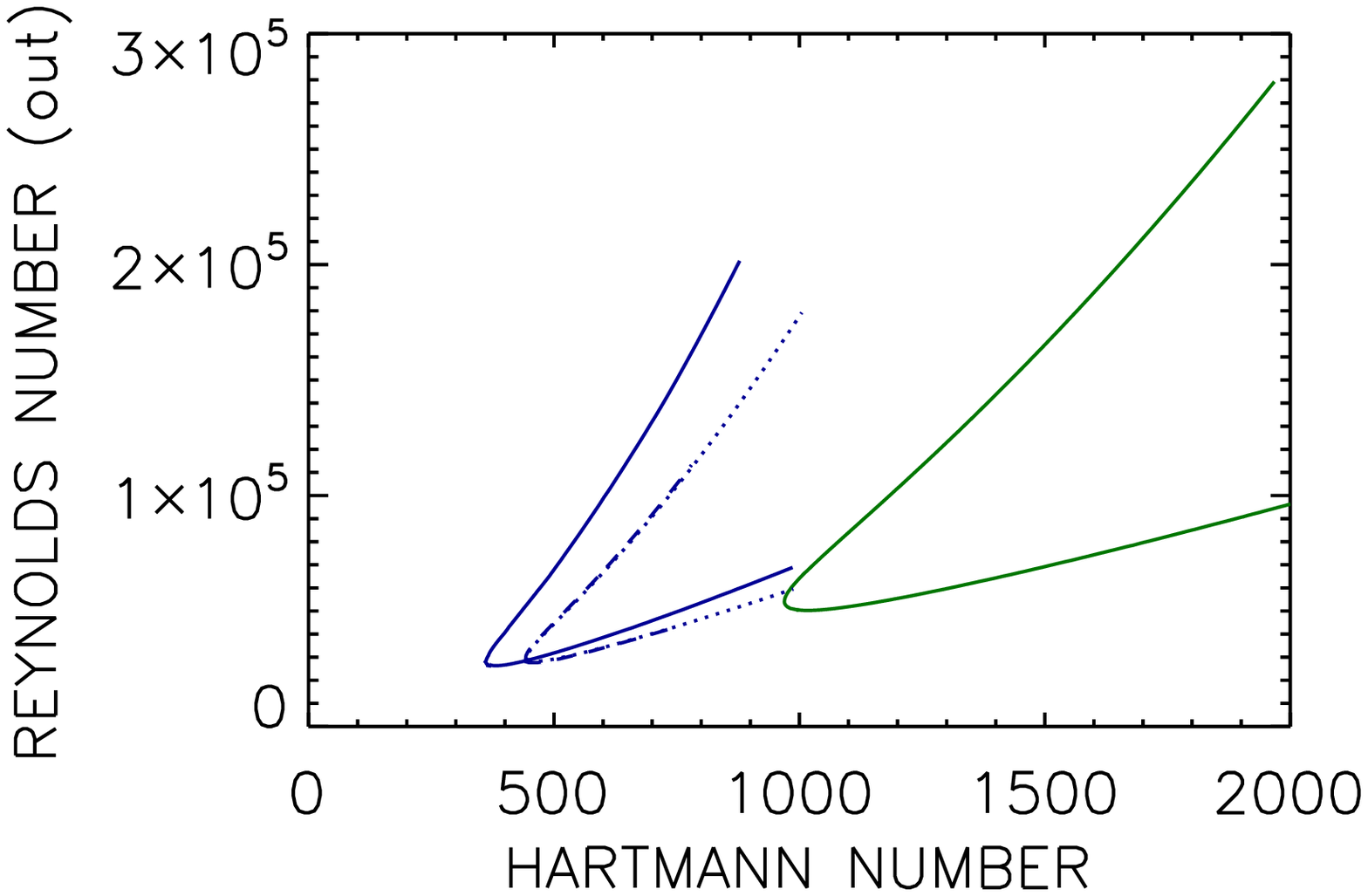}
 \includegraphics[width=0.51\textwidth]{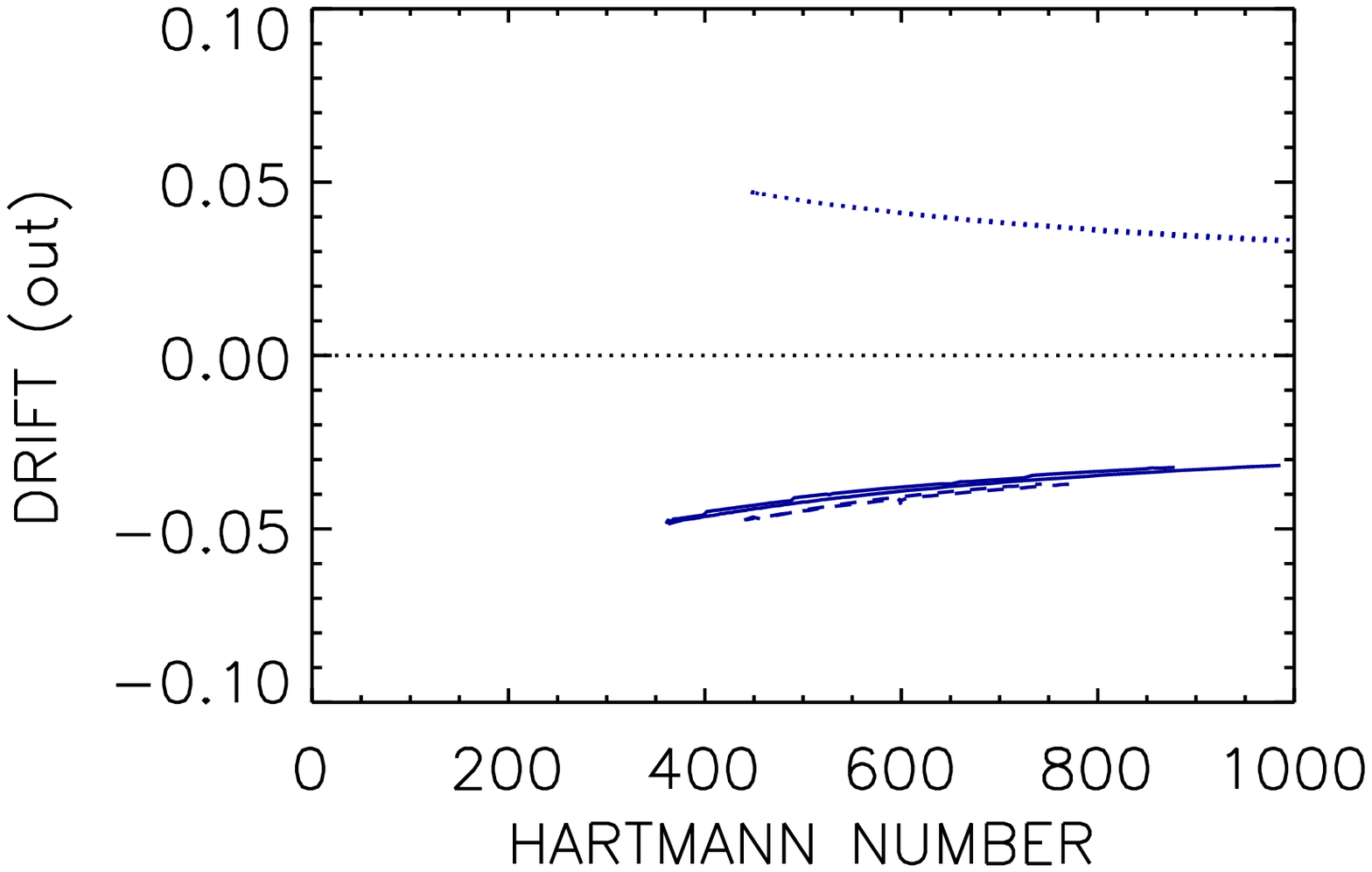}
 \caption{Left: Stability map of the non-axisymmetric modes for $\rin= 0.8$ (blue) and $\rin=0.9$ (green) of  super-rotating flow. Solid line: $m=1$, $\beta=25$. Dotted line: $m=-1$, $\beta=25$. Dashed line: $m=1$, $\beta=-25$. 
 Right panel: Drift rates. Note that for $\beta=25$ the sign of $\omdr$ differs for $m=1$ (solid) and $m=-1$ (dotted). $\mu=128$, $\Pm=1$. Insulating cylinders. } 
 \label{figm1a} 
\end{figure}

Fig.~\ref{figm1a} (left panel) gives the stability map for the non-axisymmetric modes with $m=\pm 1$, for two choices of $\rin$. The values of $\Hamin$ exceed those of the axisymmetric mode, and now $\Hamin$ also depends strongly on $\rin$. $\Hamin$ increases from 350 for the wider gap (blue line) to about 1000 for the narrower gap (green line). A general result is that the flow in the wide gap is more unstable than the flow in the narrow gap. Recall also that the Hartmann number (\ref{pm}) is defined with the weak axial field, so that the Hartmann number of the toroidal field is higher by the (large) factor $\beta$.

The critical values for $\Ha$ and $\Re$ differ only slightly for $m=\pm 1$, as do the wave numbers. The minimum Hartmann number for excitation of the mode with $m=1$ is somewhat smaller than that for $m=-1$. The drift rates, however, are significantly different, so that the phase velocities of the axial drifts also differ. The mode with $m=1$ travels upwards (in the direction of positive $z$) while the mode with $m=-1$ travels downwards (in the direction of negative $z$). The wave numbers $k R_0\approx 1$ (not shown) of the spirals are larger than the wave numbers of the axisymmetric mode, but still they are rather small so that the cells are oblong. The two non-axisymmetric modes form two different spirals. The general phase relationship is $\d z/\d \phi= -m/k$, so that the mode with positive $m$ forms a left-hand spiral while the mode with negative $m$ forms a right-hand spiral. The two spirals have slightly different excitation conditions but they travel in opposite directions.

The question is what happens with the eigensolutions for $\beta\to -\beta$. Then obviously the chirality of the background field is changed. The perturbations should react with $m\to -m$. Indeed, Fig.~\ref{figm1a} verifies that the transformation $\beta \to -\beta$ simply replaces the transformation $m \to -m$. One only finds the two possible spirals which we already know for $m=\pm 1$. The curve for $m=1$ and $\beta=-25$ in the $(\Ha/\Re)$ plane agrees with the curve for $m=-1$ and $\beta=25$. The same is true for the wave numbers, but is not true for the drift speeds, which change sign. Here only the azimuthal wave number $m$ determines the $\omdr$ value. Its values for $m=1$ and $\beta=\pm 25$ are identical (right panel of Fig.~\ref{figm1a}).

\section{The axisymmetric modes in their dependence on the inclination angle beta}\label{Axi}
\begin{figure}
\includegraphics[width=0.52\textwidth]{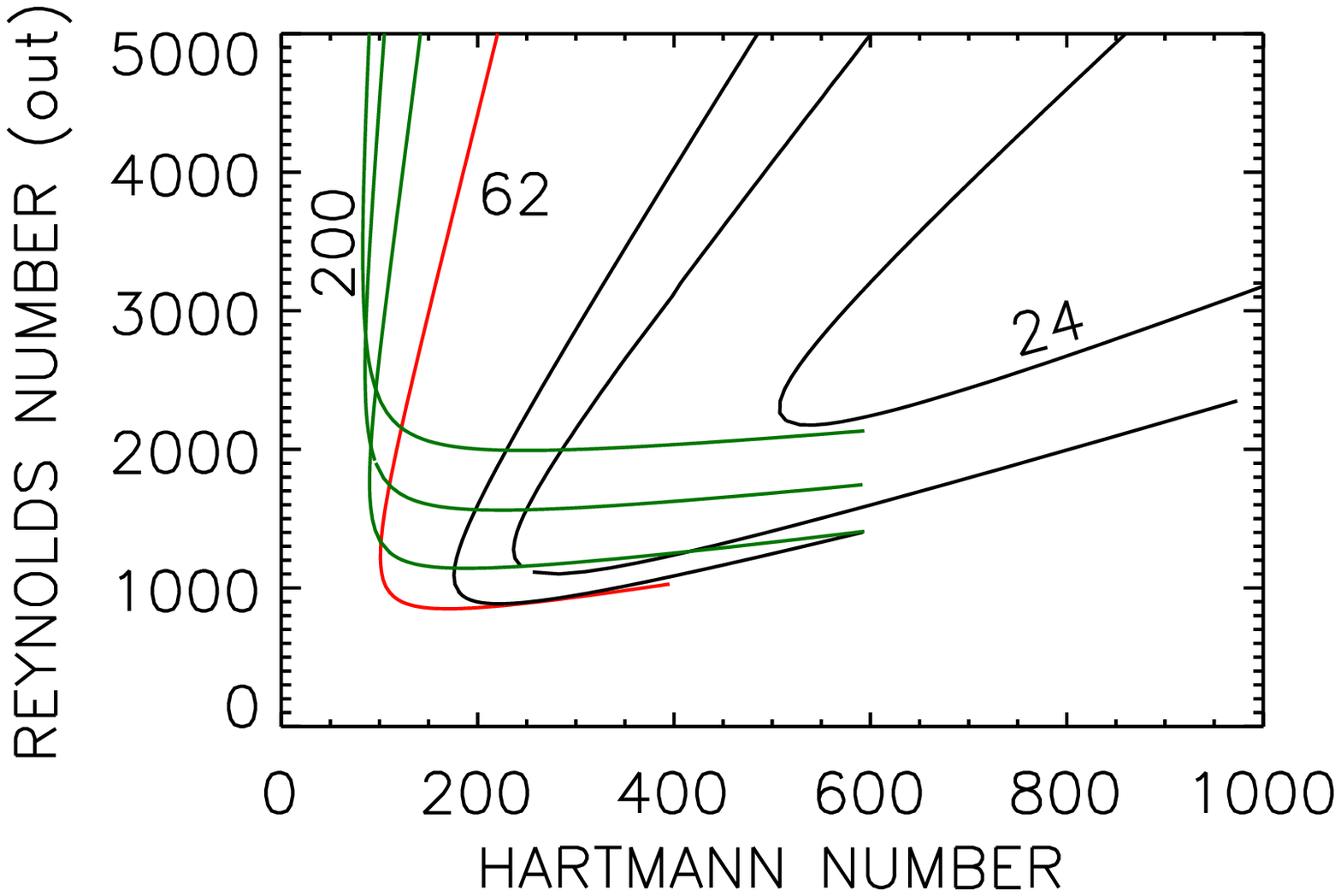}
 \includegraphics[width=0.52\textwidth]{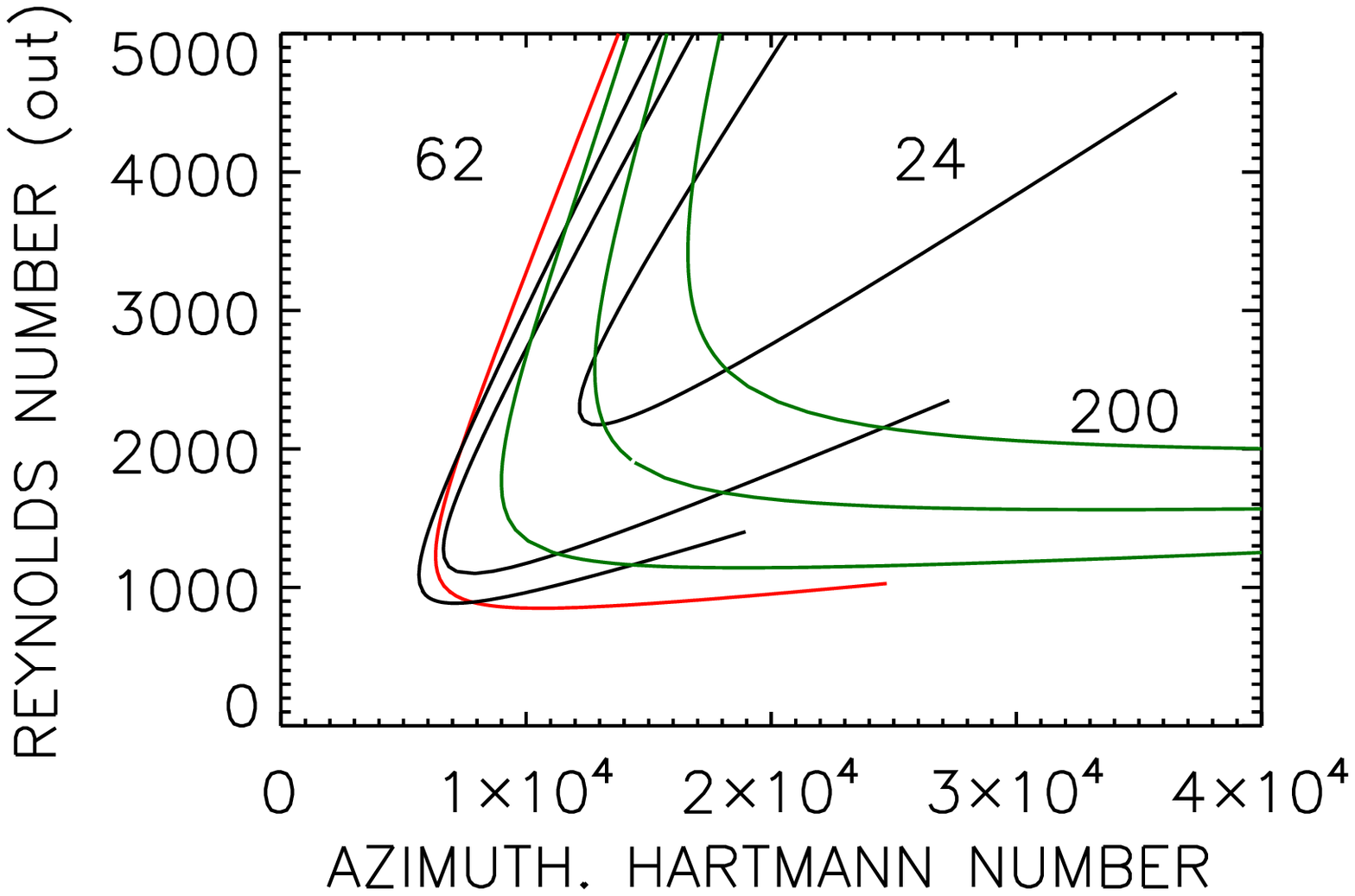}
 \caption{Stability map of axisymmetric modes for various pitch angles $\beta$. For the horizontal axis the axial Hartmann number $\Ha$ is used in the left panel, and the azimuthal Hartmann number $\Ha_\phi=\beta \Ha$ in the right panel. Black lines: $\beta<62$ (down to 24), red line: $\beta=62$, green lines: $\beta>62$ (up to 200). The lines are marked with their values of $\beta$. $\rin=0.8$, $\mu=128$, $\Pm=0.5$, $m=0$. Insulating cylinders.} 
 \label{figbeta1} 
\end{figure}

For $\beta\to 0$ the system would turn into that of the standard magnetorotational instability which, however, does not exist for super-rotation. On the other hand, for $\beta \to \infty$ the system approaches that of the super-AMRI which also does not exist for axisymmetry. Hence, there should be an optimal $\beta$ at which the instability is most easily excited, depending only on $\Pm$ for any fixed $\rin$. Figure \ref{figbeta1} shows the stability lines for $\Pm=0.5$. The horizontal axes in the two plots are the axial Hartmann number $\Ha$ (left panel) and the azimuthal Hartmann number $\Ha_\phi$ (right panel), where we recall that the azimuthal Hartmann numbers are defined by $\Ha_\phi=\beta \Ha$ formed with the azimuthal field $B_\phi$ rather than $B_z$. One finds the minimum values of $\Re$ growing for both large $\beta$ (green lines) and small $\beta$ (black lines). The red line for $\beta=62$ represents the instability with the absolutely lowest Reynolds number; all other lines are located above this line. With this low Reynolds number the azimuthal magnetic Mach number (\ref{Mm}) takes on the low value $\Mm\simeq 0.1$. It remains always constant for higher $\beta$. For $\Pm$ of order unity magnetic Mach numbers exceeding 0.1 are necessary for instability, but they must not be greater than (say) 0.3 (for $\beta \simeq 62$). The axisymmetric modes are thus not unstable for magnetic Mach numbers exceeding unity. The minimum Hartmann numbers do not depend on $\beta$ for large $\beta$ (see left panel). For low $\beta$ the minimum {\rm azimuthal} Hartmann numbers do not depend on $\beta$ (see right panel, if $\beta$ is not too small). 

Considering its right panel, Fig.~\ref{figbeta1} also demonstrates that the opening of the instability cone is largest for the optimal $\beta\simeq 62$. The cone becomes increasingly narrow for smaller $\beta$, i.e.\ for greater $B_z$ (so tending toward the standard MRI limit). A similar behaviour can be observed for much greater $\beta$. One finds the most extensive instability domain for an optimal value $\beta\simeq 60$. For smaller as well as larger values the axisymmetric instability is suppressed as the constellations with $\beta=0$ and $\beta=\infty$ are stable.

We note that the wave numbers and also the drift rates (normalized with the outer rotation rate) are small 
. The drift rates only depend slightly on $\beta$ and the Hartmann number. For small $\beta$ the wave numbers become increasingly small. In these cases the phase speed (normalized with $R_0\Omout$) is of order 0.1, while for the larger values of $\beta$ it is smaller, of order 0.01.

\section{The axisymmetric modes in their dependence on the magnetic Prandtl number}\label{Axial}
The axisymmetric modes are the solutions with the lowest critical parameter values. They shall now be considered for magnetic Prandtl number larger or smaller than unity. For $\Pm\geq 0.5$, Table \ref{tab1} gives the critical values of marginal instability of a flow with $\mu=128$ penetrated by a helical magnetic field with the optimal value $\beta=62$. These numbers may serve to model the interaction of a strongly super-rotating flow and a magnetic field with a moderate axial component. The main result is that we find the flow is unstable also for $\Pm=1$. For the models of Table \ref{tab1} the critical magnetic Reynolds numbers hardly vary. The same is true for the critical Lundquist numbers. The lines of neutral stability for $\Pm\gsim 1$ appear to scale with $\Rm$ and $\S$ rather than with $\Re$ and $\Ha$.
\begin{table}
\caption{
Eigenvalues of the axisymmetric solutions for super-rotating flows with $\mu=128$, $\rin=0.8$ and $\beta=62$. $\omega_{\rm diff}$ as in Eq.~(\ref{omdiff}). Minimal Hartmann numbers, insulating boundary conditions.
}
\label{tab1}
\centering
\begin{tabular}{lcccccccc}
\hline\hline\\

\smallskip
$\Pm$ & $\Re$ &$\Rm$&$\Hamin$ & $\S_{\rm min}$ &$\Mm$ &$ k R_0$& $\omega_{\rm dr}$&$\omega^{\rm R}/\omega_{\rm diff}$\\ \\
\hline
\\
0.5 &1233&617 &101&51&0.19 &0.54&-0.015&-9.3\\
1 & 544 &544& 38.3 &38& 0.22&0.98 &- 0.027&-14.7\\
 2& 294&588&17.4& 35& 0.27 &1.57 &-0.039 &-22.9\\
 3 & 218 &654 &11.3& 34&0.30 & 2.04 &-0.044&-25.7 \\ \\

 \hline
\end{tabular}
\end{table} 

The negative sign of the drift rates (\ref{omdr}) for positive $\beta$ is another important result. As we shall demonstrate below, this sign is opposite to that for flows with sub-rotation. The instability pattern thus migrates ``pole-wards'' (i.e. in positive $z$-direction) along the rotation axis for super-rotation and ``equator-wards'' (i.e. in negative $z$-direction) for sub-rotation for all positive $\beta$. As noted above the axial migration is slower than the rotation time scale faster than the magnetic diffusion time scale.
\begin{figure}
 \includegraphics[width=0.51\textwidth]{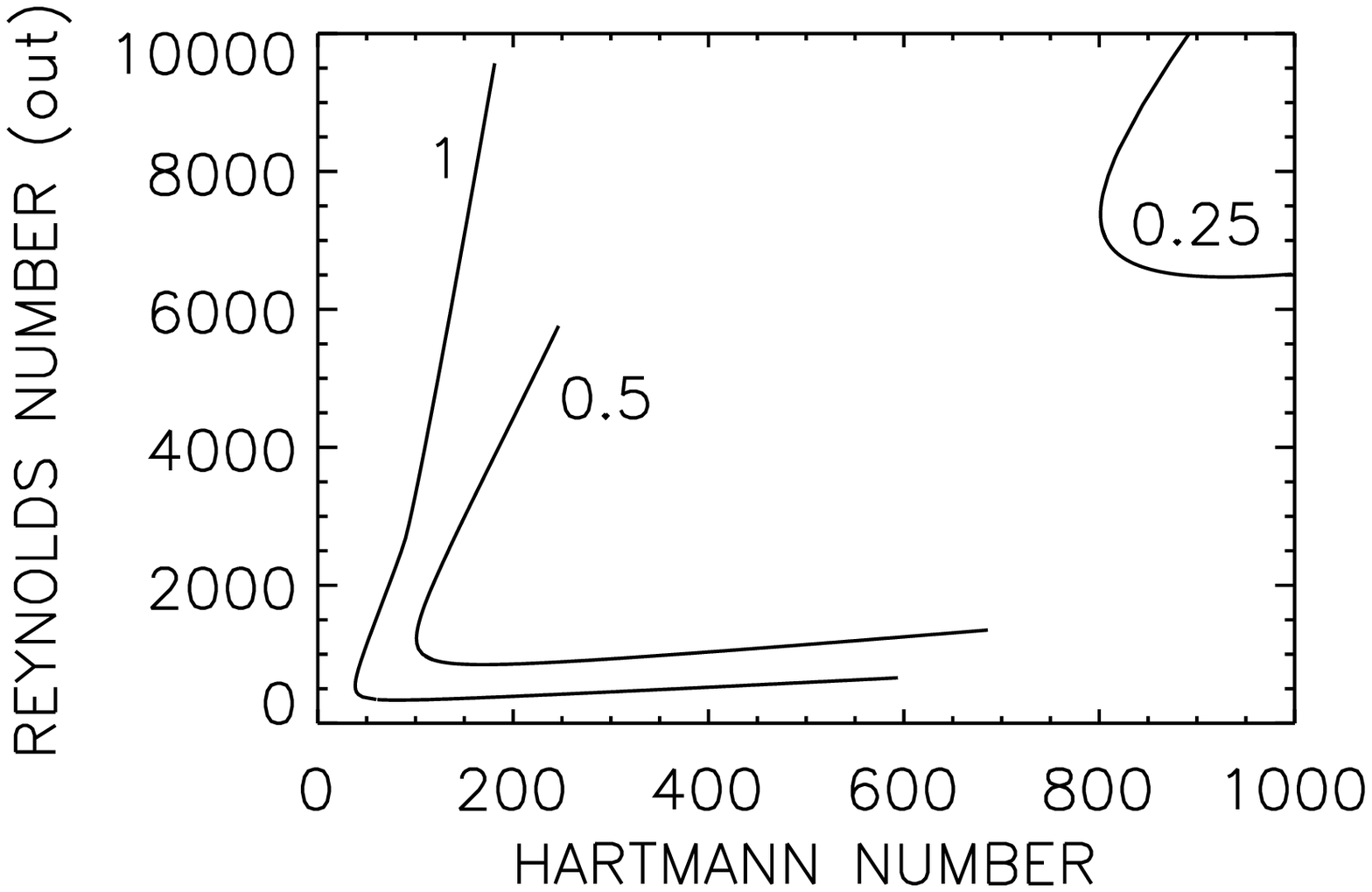}
 \includegraphics[width=0.51\textwidth]{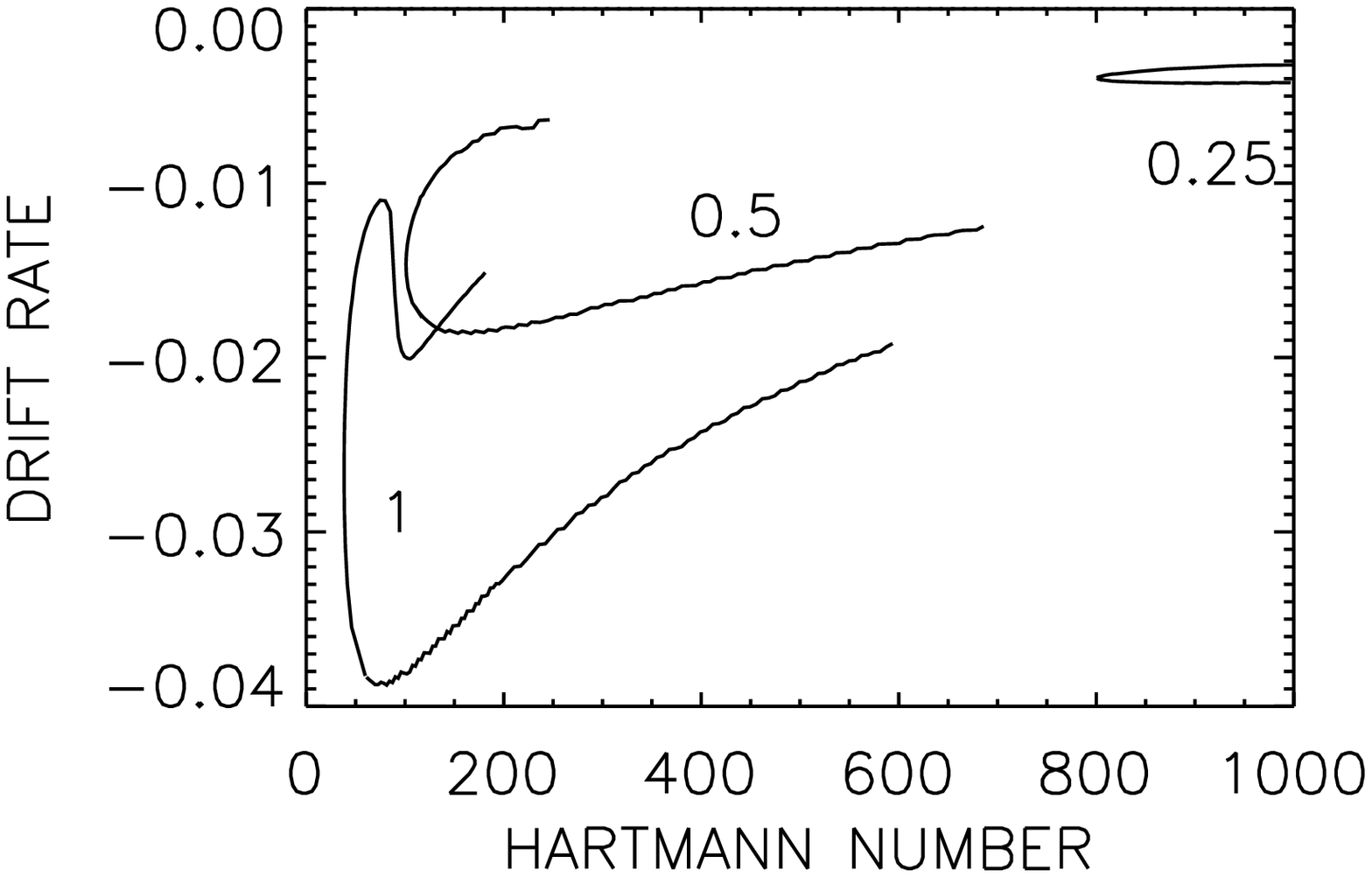}
 \caption{Left panel: Stability map of the axisymmetric  modes of super-rotating flows with $\mu=128$. 
 Right panel: Normalized drift frequencies. The curves are marked with their magnetic Prandtl numbers. $\beta=62$, $m=0$, $\rin=0.8$. Insulating cylinders. } 
 \label{fig2p} 
\end{figure}
\begin{figure}
\includegraphics[width=0.48\textwidth]{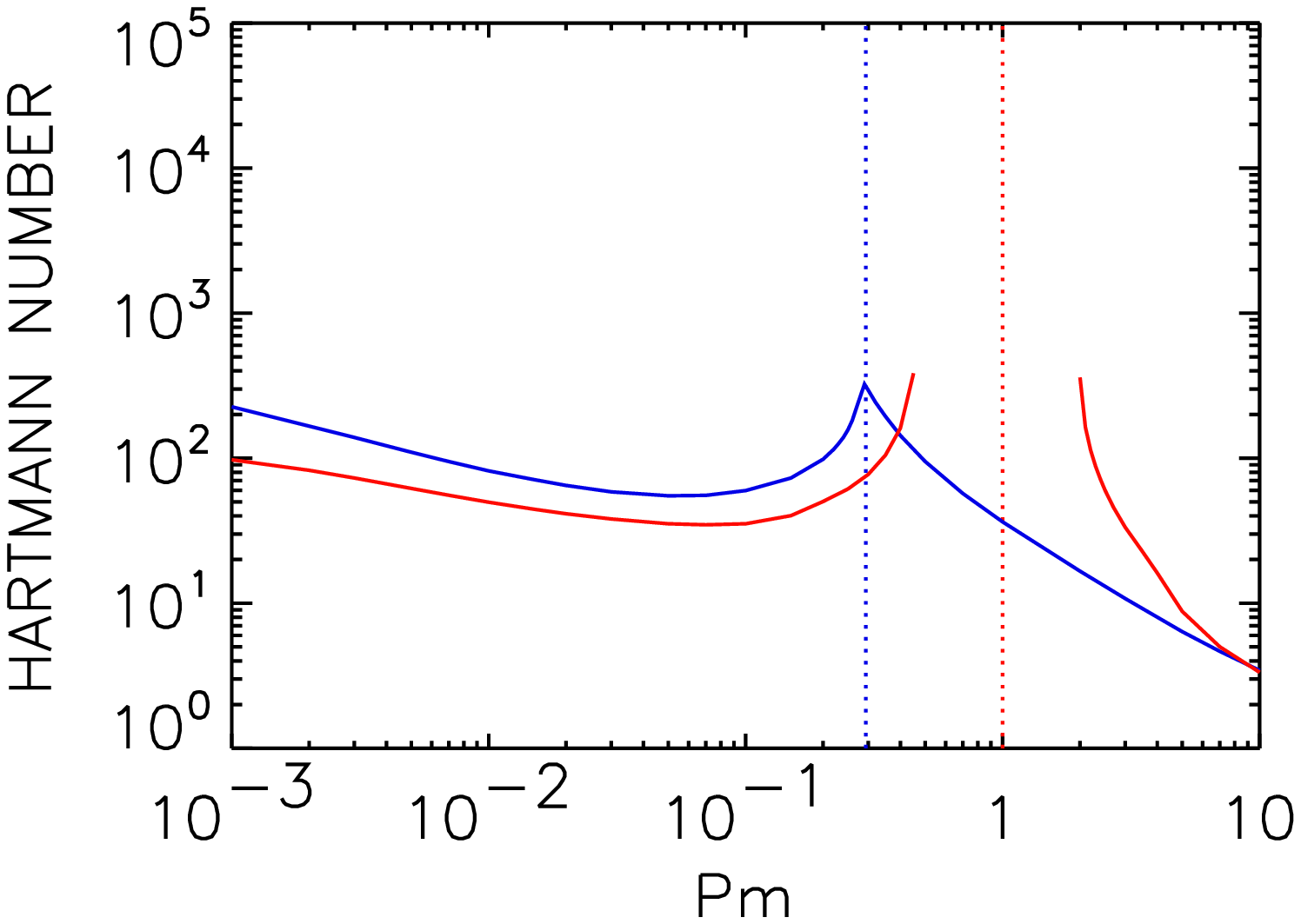}
 \includegraphics[width=0.48\textwidth]{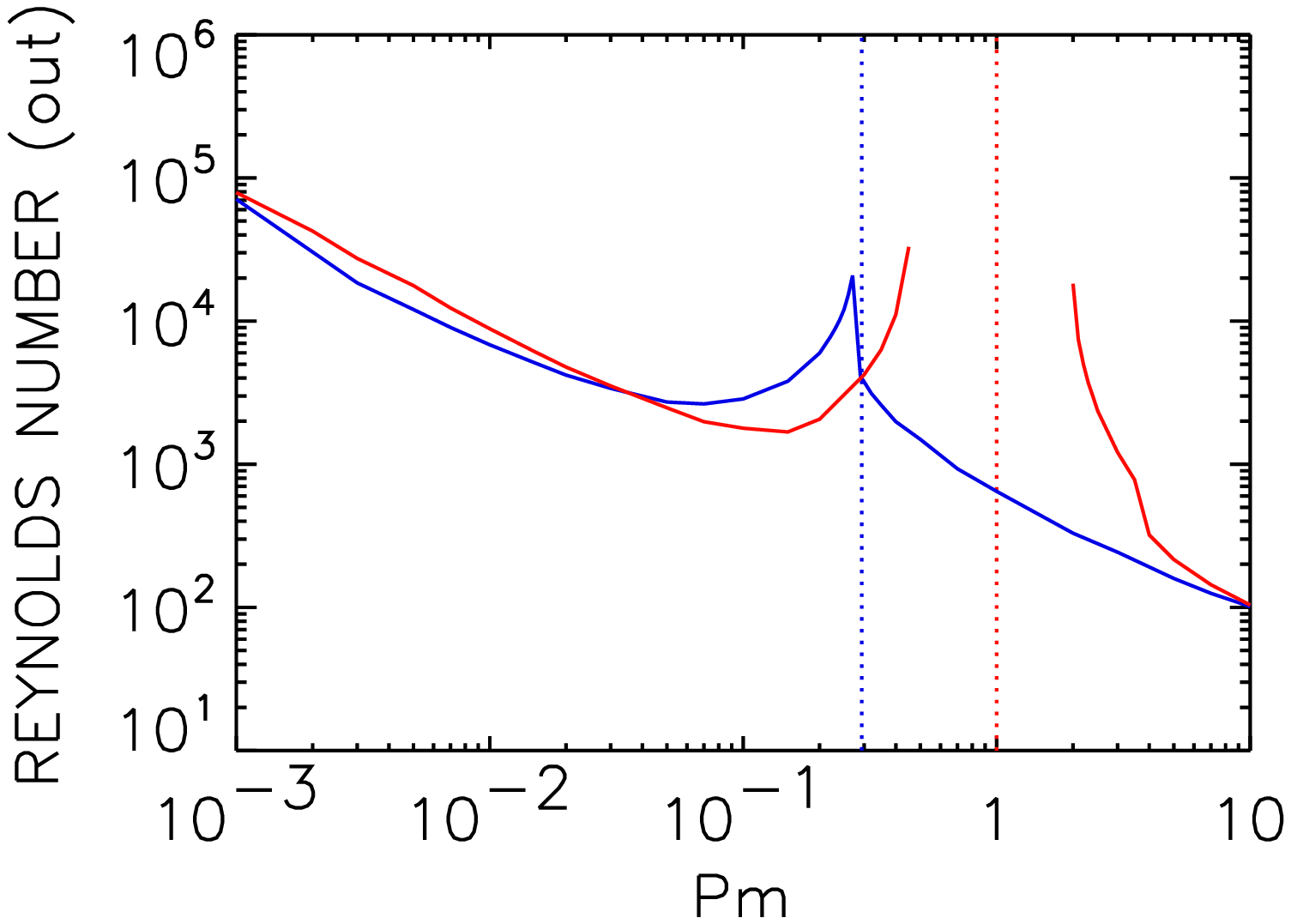}
 \caption{The minimal Hartmann number $\Hamin$ (left panel) and the related Reynolds numbers (right panel) for neutral  stability as function of the magnetic Prandtl number for two different boundary conditions (red lines: perfectly conducting walls, blue lines: insulating walls). The two vertical dotted lines mark $\Pm=0.28$ and $\Pm=1$,  resp. $\mu=128$, $\beta=50$, $m=0$, $\rin=0.8$.}
 \label{fig2q}
\end{figure}

The stability maps for $\Pm\leq 1$ with insulating boundary conditions are shown in Fig.~\ref{fig2p}. They all have the same typical conical structure  as in Fig.~\ref{figm0a}. For supercritical Hartmann numbers there are always two Reynolds numbers between which the flow is unstable. The slopes $\d \Re/\d \Ha$ of both branches are again positive and very similar. There is always a minimum Hartmann number $\Ha_{\rm min}$ at $\d \Re/\d \Ha=\infty$ below which the flow is stable. We note that this `oblique-cone' geometry of the instability domain previously only appeared for the non-axisymmetric modes of MRI and AMRI. The ``helical'' magnetorotational instability (HMRI) with super-rotation (``super-HMRI'') is the only magnetic instability known sofar where rapid rotation stabilizes the {\em axisymmetric} mode. Rotation excites the instability, but it can also be too fast for its existence. This is quite opposite to the excitation conditions of the axisymmetric (or channel) modes of standard MRI \citep{GR12} or HMRI with negative shear \citep{SG06,RGH18} which do not possess upper limits of the Reynolds number. The HMRI with super-rotation (``super-HMRI'') is thus {\em much} more {stable} than HMRI with sub-rotation.

The two branches of the curves in the left panel of Fig.~\ref{fig2p} limit the magnetic Mach number (\ref{Mm}) of the azimuthal field
to the small value of O(0.1) for the unstable modes. Flows with higher magnetic Mach numbers, i.e.\ with faster rotation or weaker field, are stable. Instability occurs for azimuthal Mach numbers only between 0.05 and 0.1.

As also indicated by the $7^{\rm th}$ column of Table 1 one finds reduced wave numbers $k$ for the unstable modes with $\Pm<1$, so that the vortices become extremely long in the axial direction. As an immediate consequence, the radial components of flow and field become smaller and smaller.
Such small wave numbers also make the problem increasingly difficult numerically. 

The right panel of Fig.~\ref{fig2p} again shows the eigensolutions possessing very small values of the drift rates if normalized with the outer rotation rate. We note, however, that with $\omega^{\rm R}/\omega_{\rm diff}\gsim 10$, the mode migrates much faster in the axial direction than the diffusion time scales indicate.

Figure \ref{fig2q} shows the minimum Hartmann numbers and corresponding Reynolds numbers over a broad range of $\Pm$ values, and including also results for insulating and conducting boundaries. For both $\Pm\ll1$ and $\Pm\gg1$ the two boundary conditions yield broadly similar results, but for $\Pm=O(1)$ there are significant differences. For conducting boundaries there are two completely separate branches, one existing for $\Pm\le0.45$ and the other for $\Pm\ge2$. No instabilities were found in the intermediate range $\Pm\in(0.45,2)$. Since this gap includes $\Pm=1$, this strongly suggests a double-diffusive type of instability. However, if we consider the insulating boundaries, there are also two separate branches, but now there is no gap in between them. Instead, around $\Pm\approx0.28$ there is simply a transition from one branch to the other, that is, a mode crossing regarding which instability has the lowest $\Ha_{\rm min}$ value. There are no values of $\Pm$ though for which no instabilities exist. Also, there is now nothing special in the vicinity of $\Pm=1$, which is seen to be just a part of the `large-$\Pm$' branch. These modes are therefore clearly different from a strict double-diffusive instability, and merely have the qualitative feature that both modes `prefer' to be away from the cross-over point at $\Pm\approx0.28$.


The phase velocity along the $z$-coordinate of the travelling axisymmetric mode is
\begin{eqnarray}
\frac{\d z}{\d t}= - \frac{\omdr}{kR_0} R_0\Om_{\rm max},
 \label{drift1} 
\end{eqnarray}
where $\omdr$ and $k R_0$ are the normalized frequencies and wave numbers. Hence, negative $\omdr$ values as given in Table \ref{tab1} for positive $\beta$ describe a wave pattern drifting in the positive-$z$ direction (``pole-ward''). The axial phase velocity of the models of Table \ref{tab1} is $0.01\ldots0.04$ in units of $R_0 \Omout$. The drift frequency is thus much lower than the rotation rate. On the other hand, as the last row in Table \ref{tab1} shows, it is faster than diffusion by one order of magnitude.


The question arises how the solutions behave under the transformation 
$\beta\to -\beta$. 
The curves for the Reynolds numbers and the wave numbers for neutral stability are identical in both cases. The drift rates, however, are different, always satisfying $\beta \omdr>0$, that is, $\omdr$ and $\beta$ always have the same sign. 
The correctness of this statement has empirically been proven by the MHD experiment PROMISE where an axially migrating axisymmetric perturbation pattern could be excited for the proper combination of a rotation law with negative shear and a helical magnetic field. The migration direction changes with the change of the sign of $\beta$ \citep{SS12}. 
 One can certainly imagine that fields with the opposite sign of chirality generate the opposite sign of (\ref{drift1}). Also numerical simulations demonstrate the invariance of the 
 solutions with neutral instability as invariant 
 against the simultaneous transformation $\beta\to -\beta$ and $\omdr\to -\omdr$. If the wave-like solution with a certain $\beta$ travels (say) upward (in direction $+z$) then another solution exists for $-\beta$ travelling in the opposite direction. Models in Table \ref{tab1} with negative $\beta$, therefore are travelling in negative $z$-direction, as also the models in Table \ref{tab2} do with positive $\beta$. In this Table for a demonstration also the 
numbers are given for $\beta=\pm 2$ where indeed only the sign of $\omdr$ is changed while the Hartmann/Reynolds numbers remain unaltered.

\section{Phase relations}
One may ask how the flow and field patterns that migrate along the $z$-axis relate to one another. Is there a shift between the maxima of flow and field and, if yes, what is its dependence on the background field or the rotation law? For several well-defined models we shall present the phase relations between the azimuthal flow perturbations and the azimuthal field perturbations for $m=0$. For super-rotating Taylor-Couette flows four cases with $0.5 \leq \Pm \leq 3$ are considered, whose critical values are given in Table \ref{tab1}. For comparison, we made similar calculations for a set of sub-rotating flows with various values of $\beta$ and a fixed magnetic Prandtl number $\Pm=0.1$ (Table \ref{tab2}).

From
\begin{eqnarray}
 b_\phi=(b^{\rm R} + \i\ b^{\rm I}) {\rm e}^{\i\psi}, \ \ \ \ \ \ \ \ \ \ \ \ u_\phi=(u^{\rm R} + \i\ u^{\rm I}) {\rm e}^{\i\psi},
 \label{bu} 
\end{eqnarray}
where the superscripts R and I denote the real and imaginary parts of the quantities, one obtains for the vertical waves of the azimuthal perturbations $b_\phi$ and $u_\phi$
\begin{eqnarray}
 b_\phi=b^{\rm R} \cos\psi - b^{\rm I} \sin\psi = b \sin(\psi-\delta_b),\nonumber\\
 u_\phi=u^{\rm R} \cos\psi - u^{\rm I} \sin\psi = u \sin(\psi-\delta_u)
 \label{b} 
\end{eqnarray}
with $\psi$ as the actual phase and the $\delta$'s as their phase shifts. Then
\begin{eqnarray}
\delta_b= \arctan\frac{b^{\rm R}}{b^{\rm I}}, \ \ \ \ \ \ \ \ \delta_u= \arctan\frac{u^{\rm R}}{u^{\rm I}}.
 \label{delta} 
\end{eqnarray}
We are only interested in the phase difference $\delta=\delta_b-\delta_u$. In order to exclude the influence of the boundary conditions  we consider this quantity only in a central region between inner and outer radius. The two waves are in phase if $\delta\simeq 0$ there. If the phase differences are given in degrees, then for $\delta\simeq 90^\circ$ the waves are out of phase.

\begin{figure}
 \centering
\vbox{
\hbox{ \includegraphics[width=0.33\textwidth]{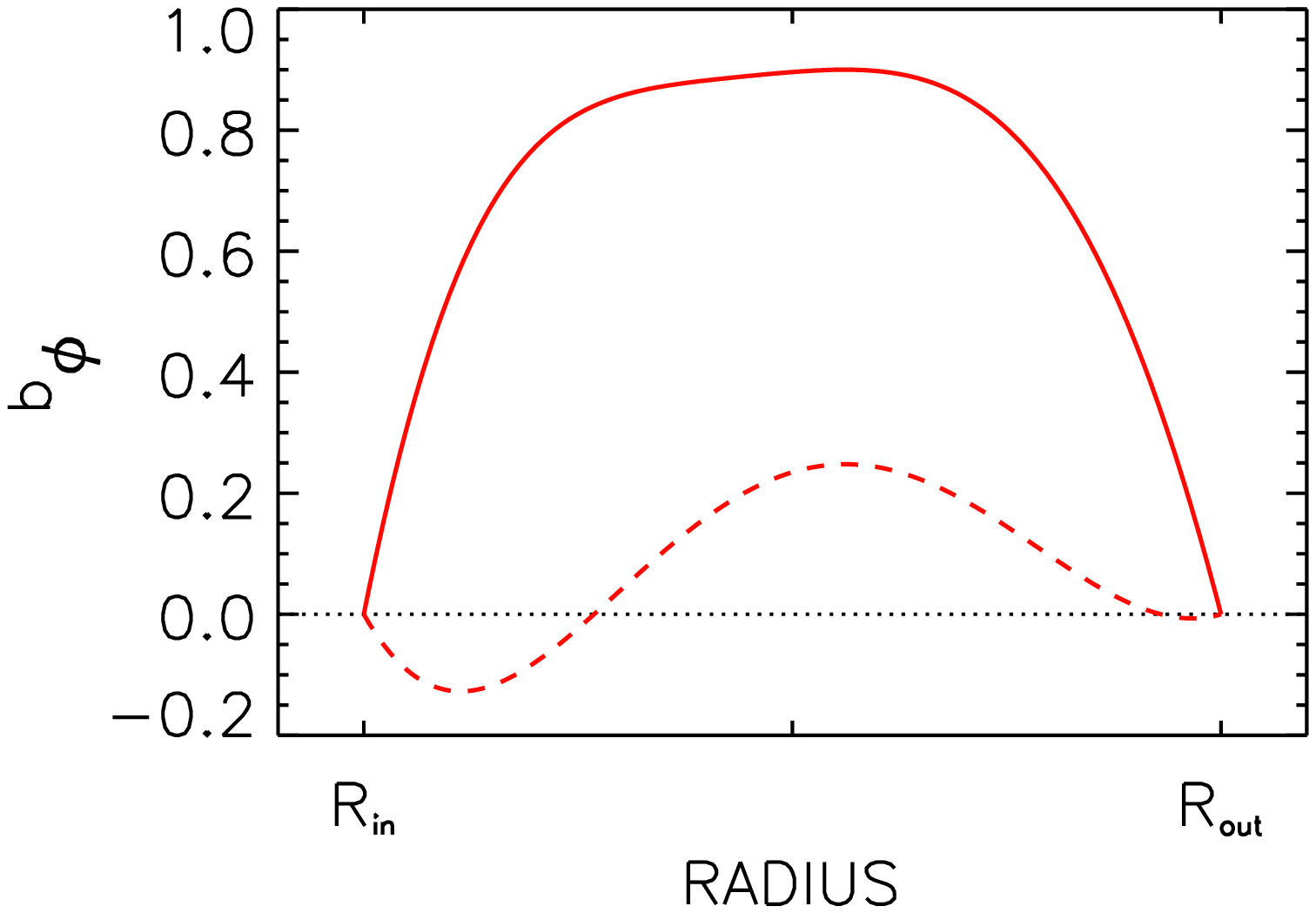}
 \includegraphics[width=0.33\textwidth]{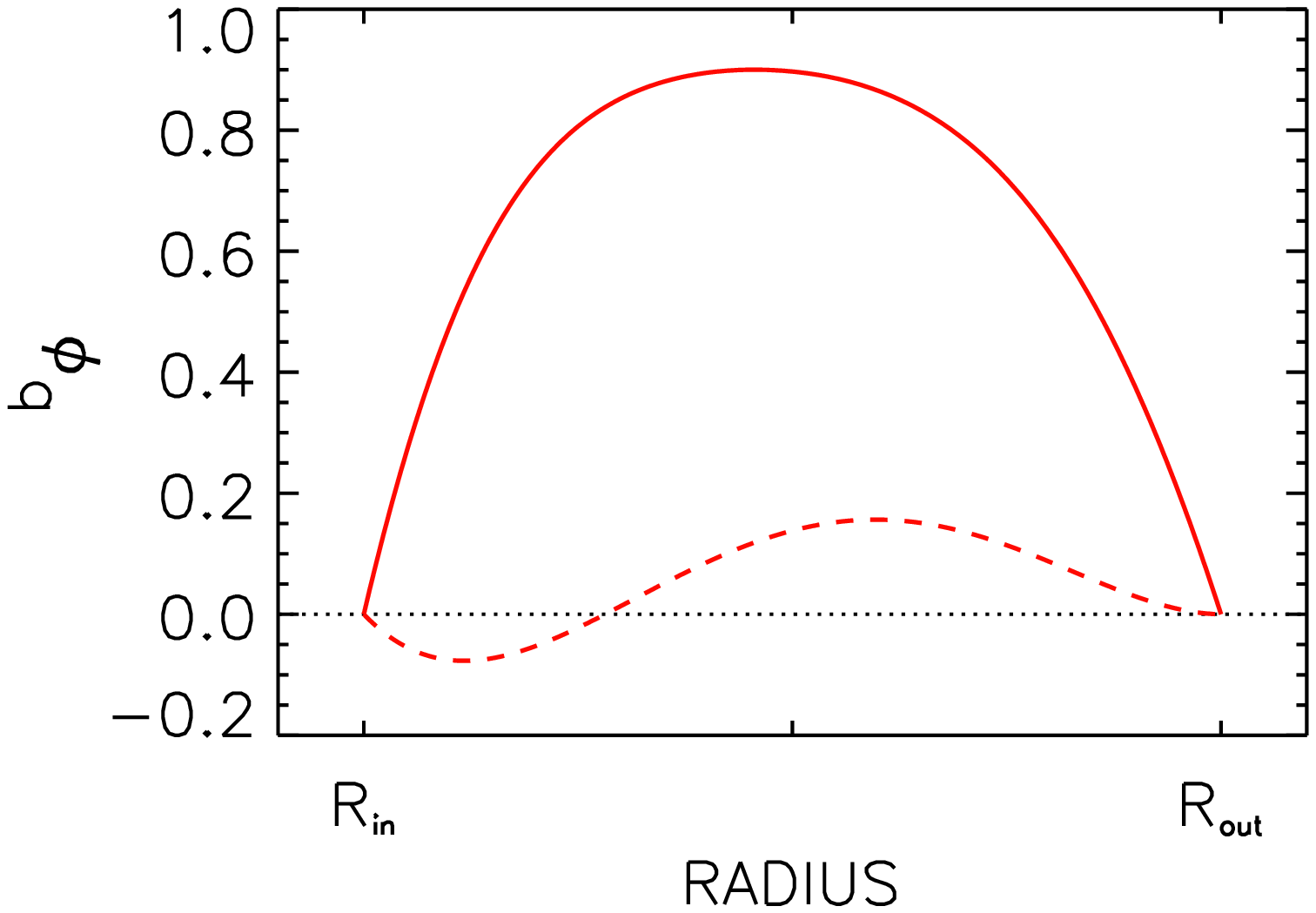}
\includegraphics[width=0.33\textwidth]{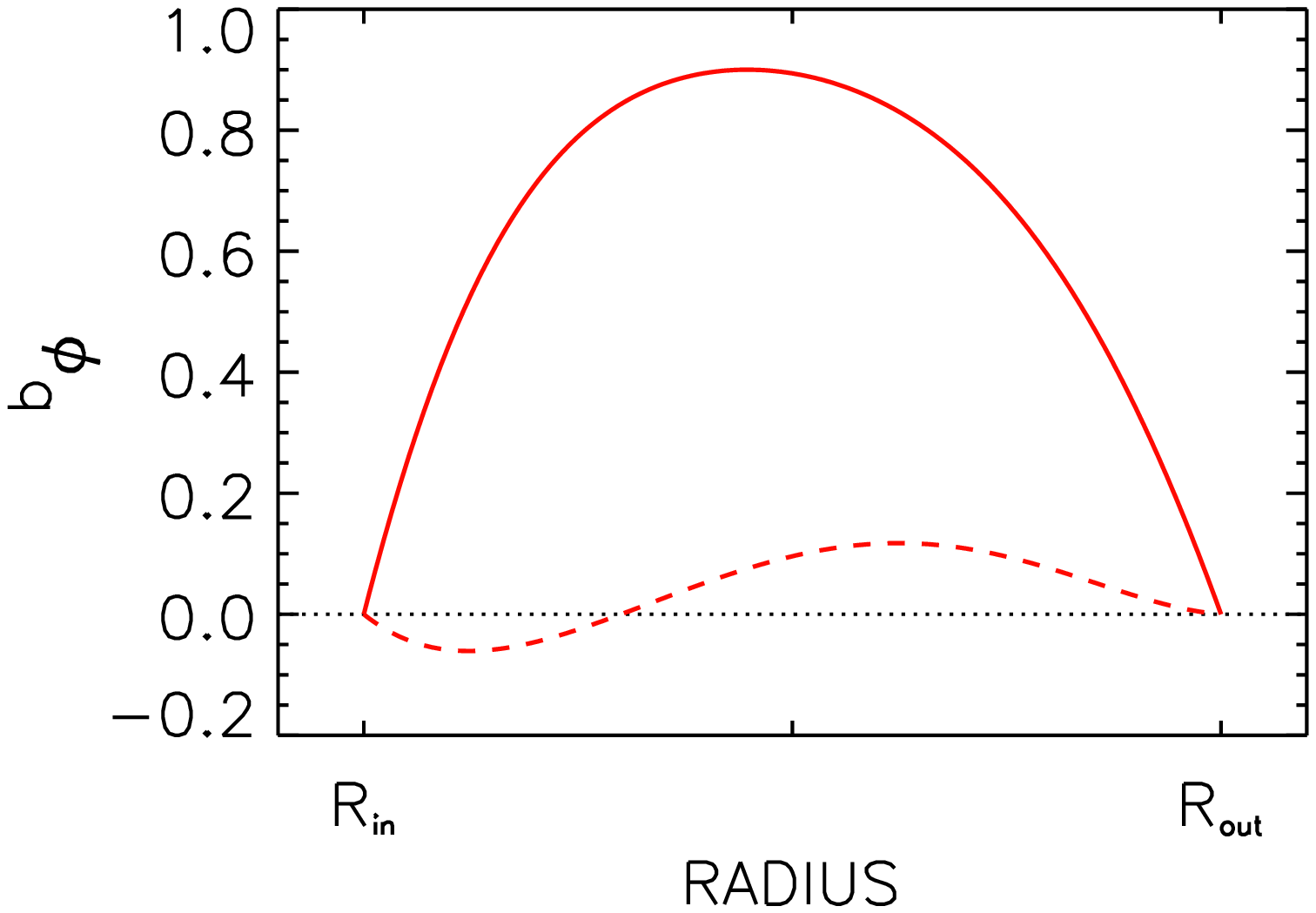} }
 \hbox{ \includegraphics[width=0.33\textwidth]{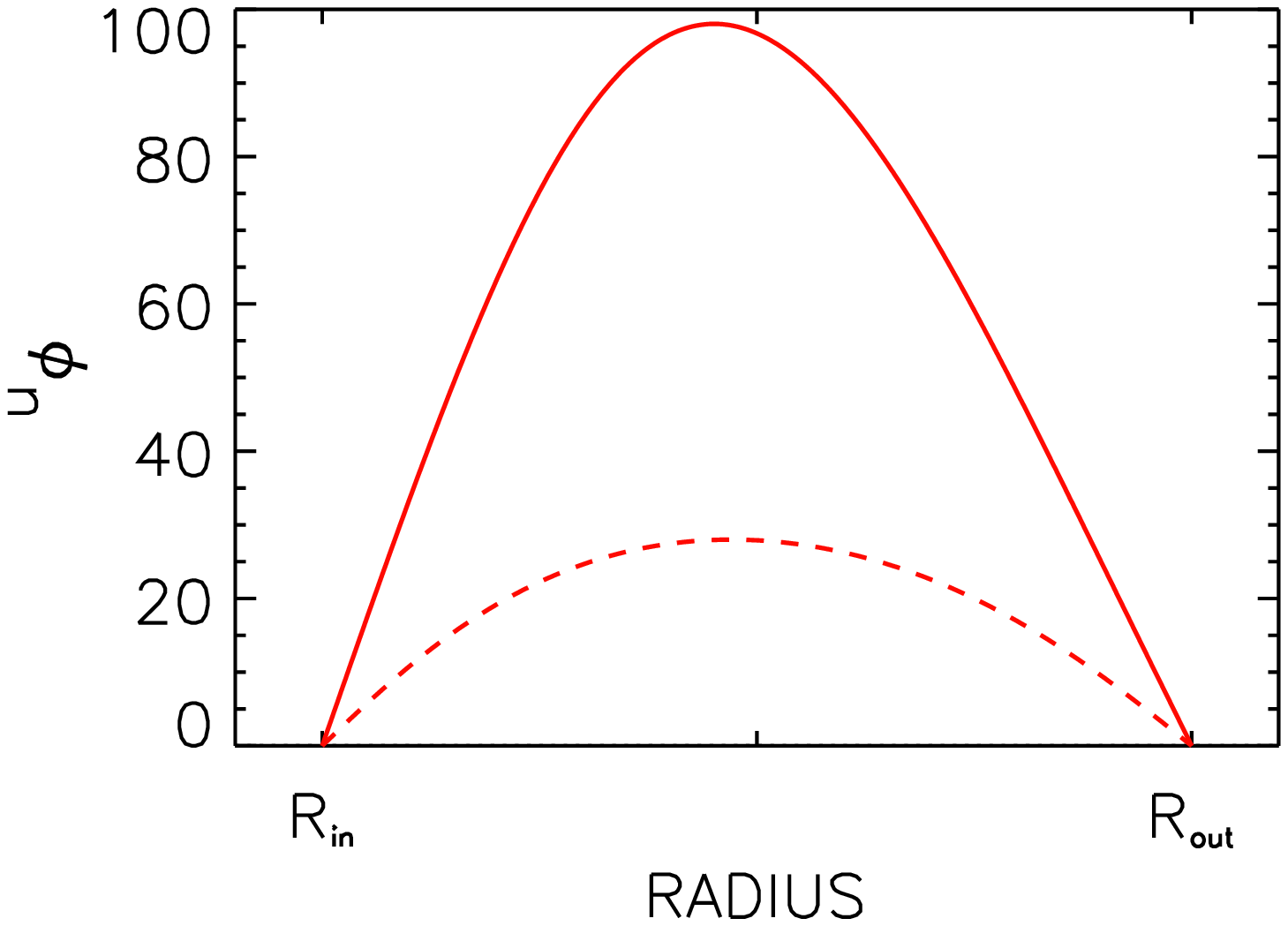}
 \includegraphics[width=0.33\textwidth]{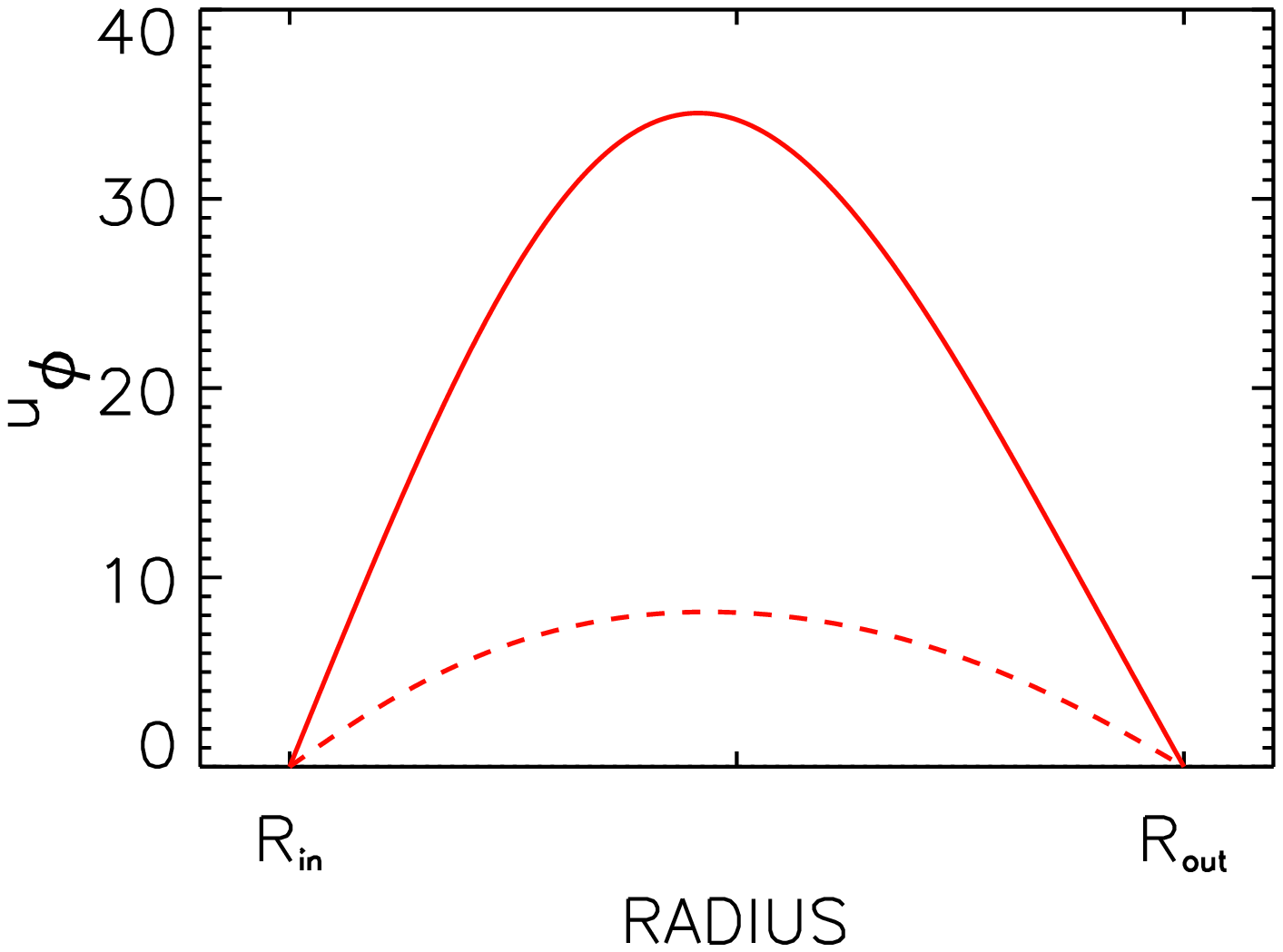}
 \includegraphics[width=0.33\textwidth]{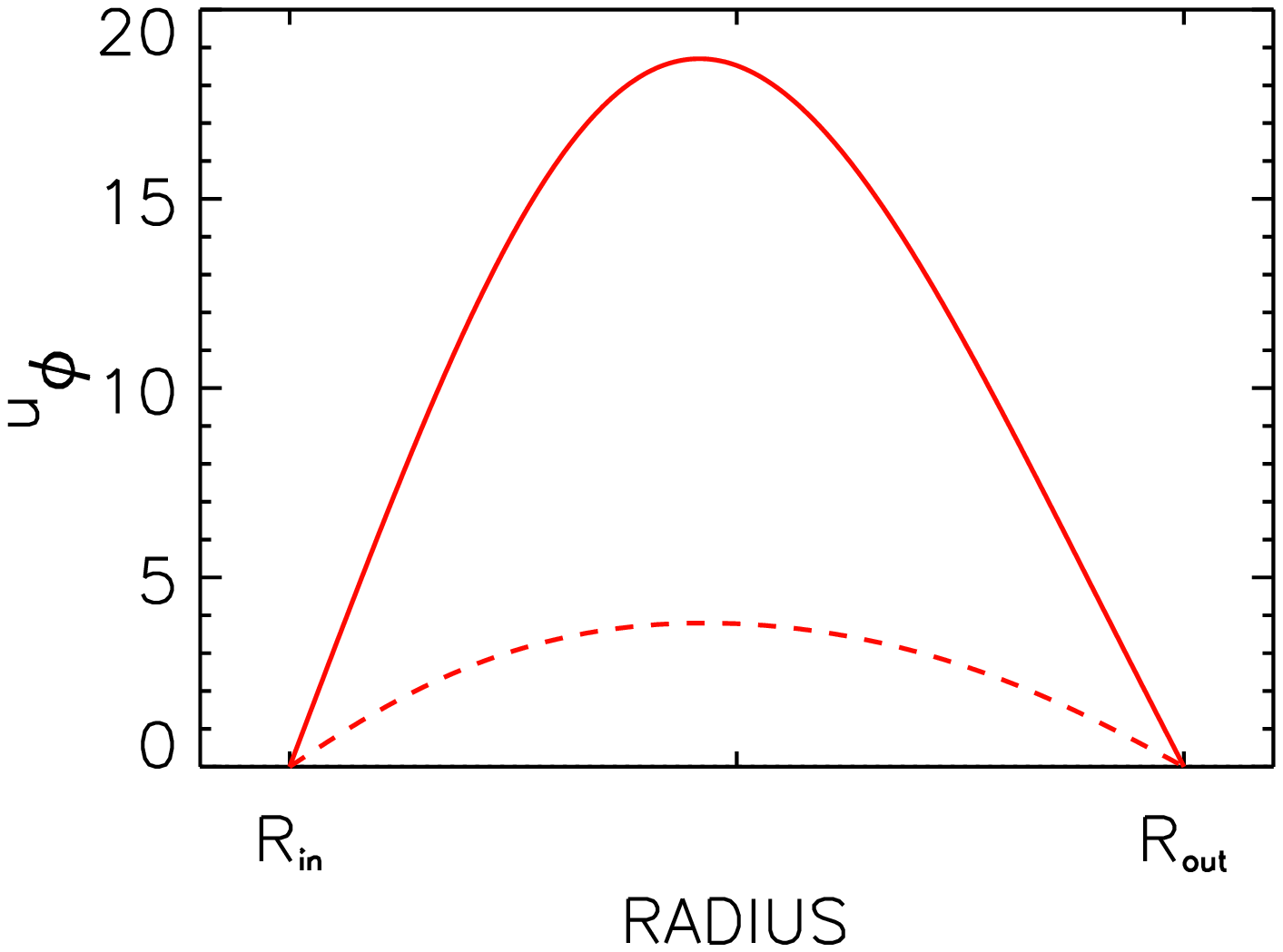}}
 }
 \caption{Radial eigenfunctions for various $\Pm$ of the azimuthal components $b_\phi$ (top) and $u_\phi$ (bottom)  with $\mu=128$ (super-rotation). Solid lines: real parts, dashed lines: imaginary parts. The functions   $b_\phi$  and $u_\phi$ of a model contain a common arbitrary factor. From left to right: $\Pm=1$, $\Pm=2$, $\Pm=3$. $\beta=62$. The critical parameters  of the models are given in Table \ref{tab1}. $m=0$, $\rin=0.8$. Insulating cylinders.} 
 \label{fig3a} 
\end{figure}

\begin{figure}
 \centering
\vbox{
\hbox{ \includegraphics[width=0.33\textwidth]{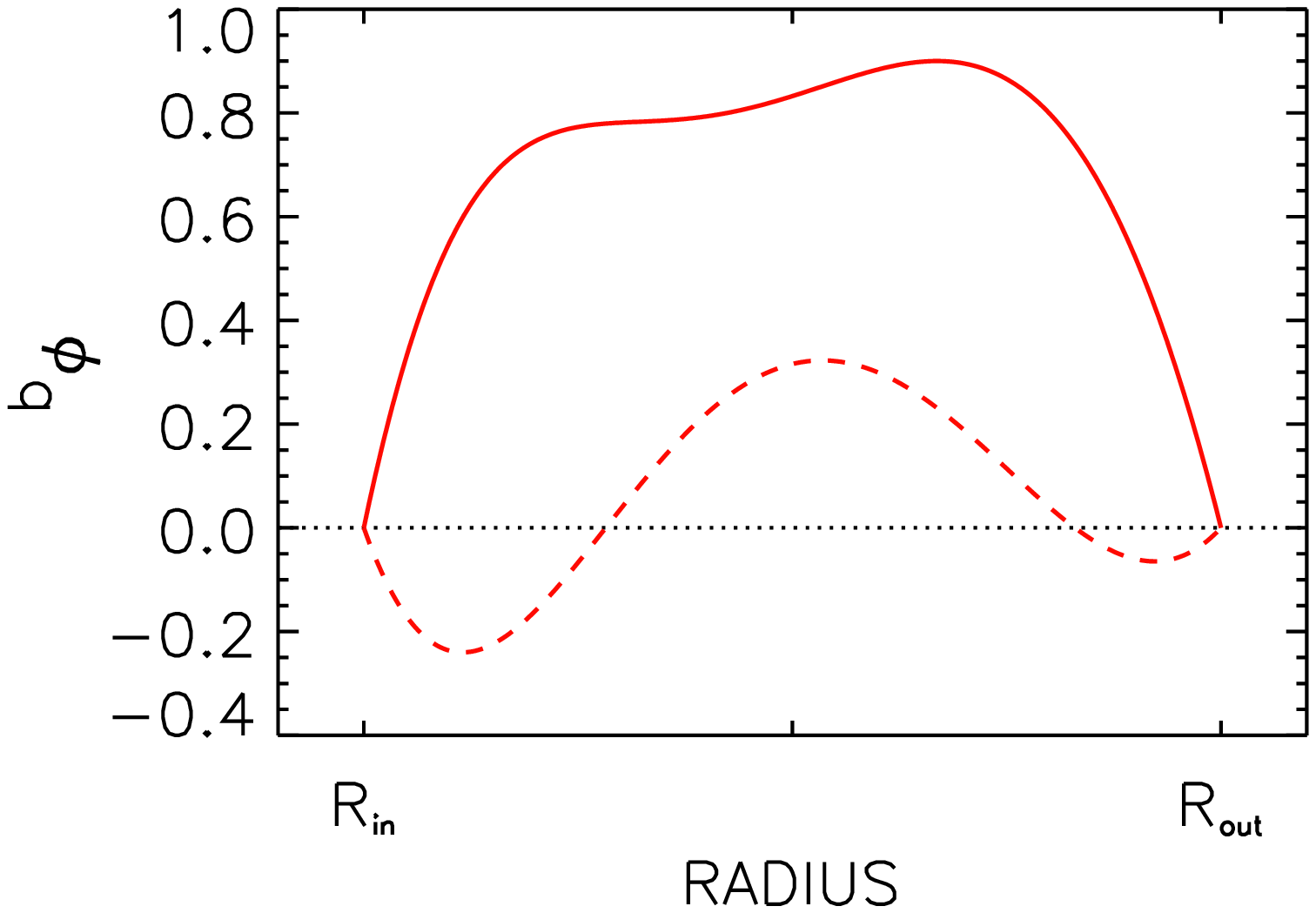}
 \includegraphics[width=0.33\textwidth]{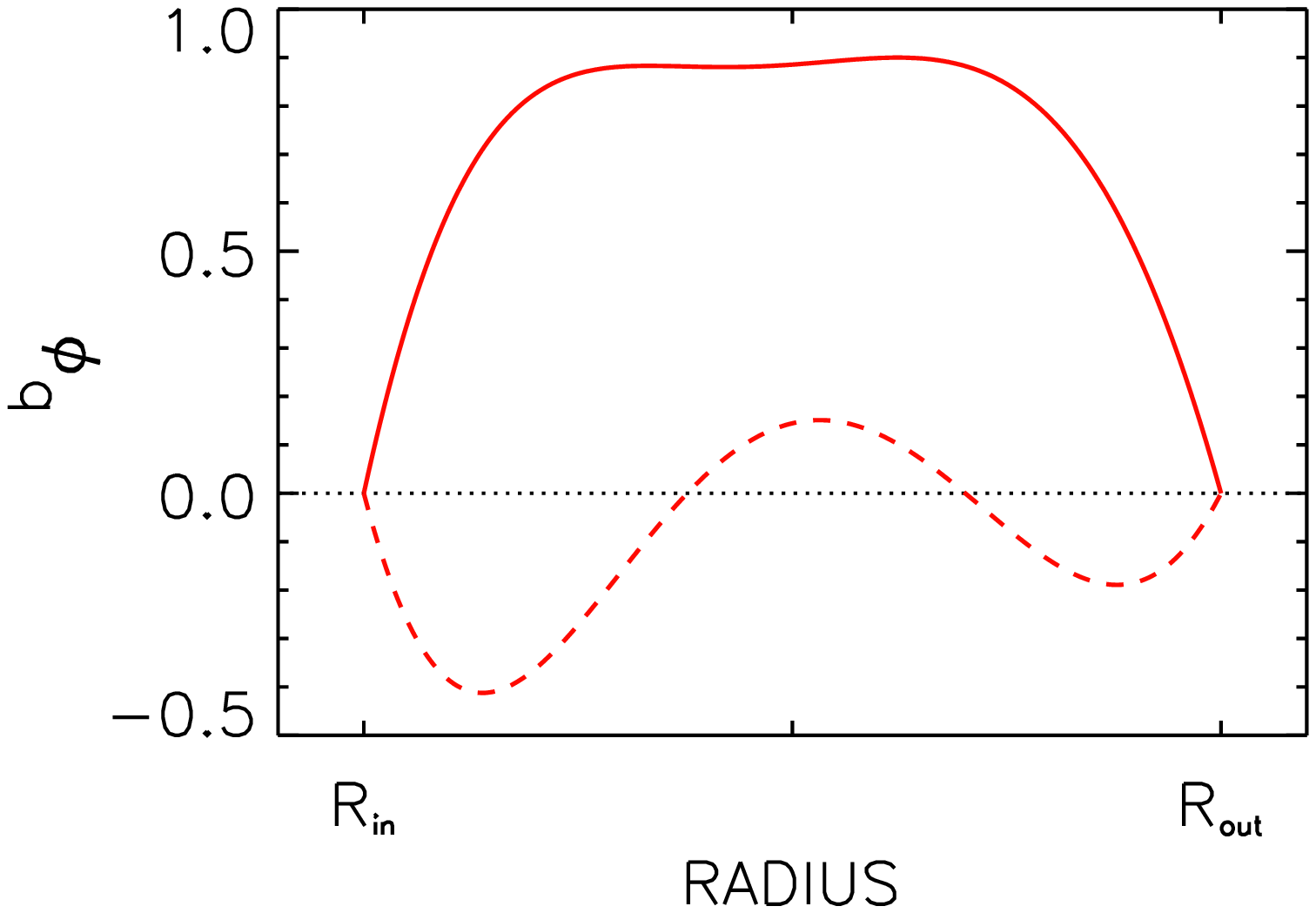}
 \includegraphics[width=0.33\textwidth]{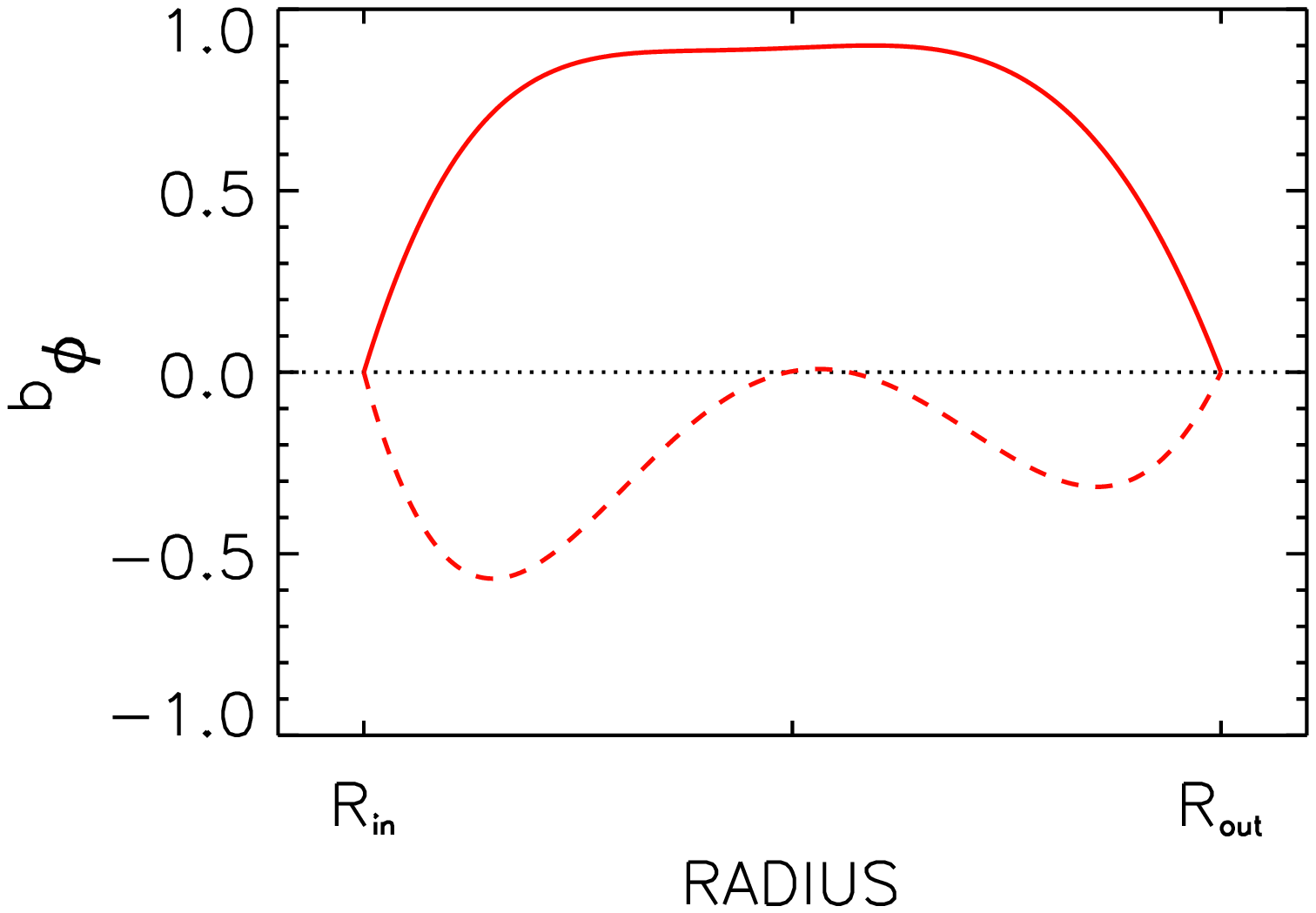}}
 \hbox{
 \includegraphics[width=0.33\textwidth]{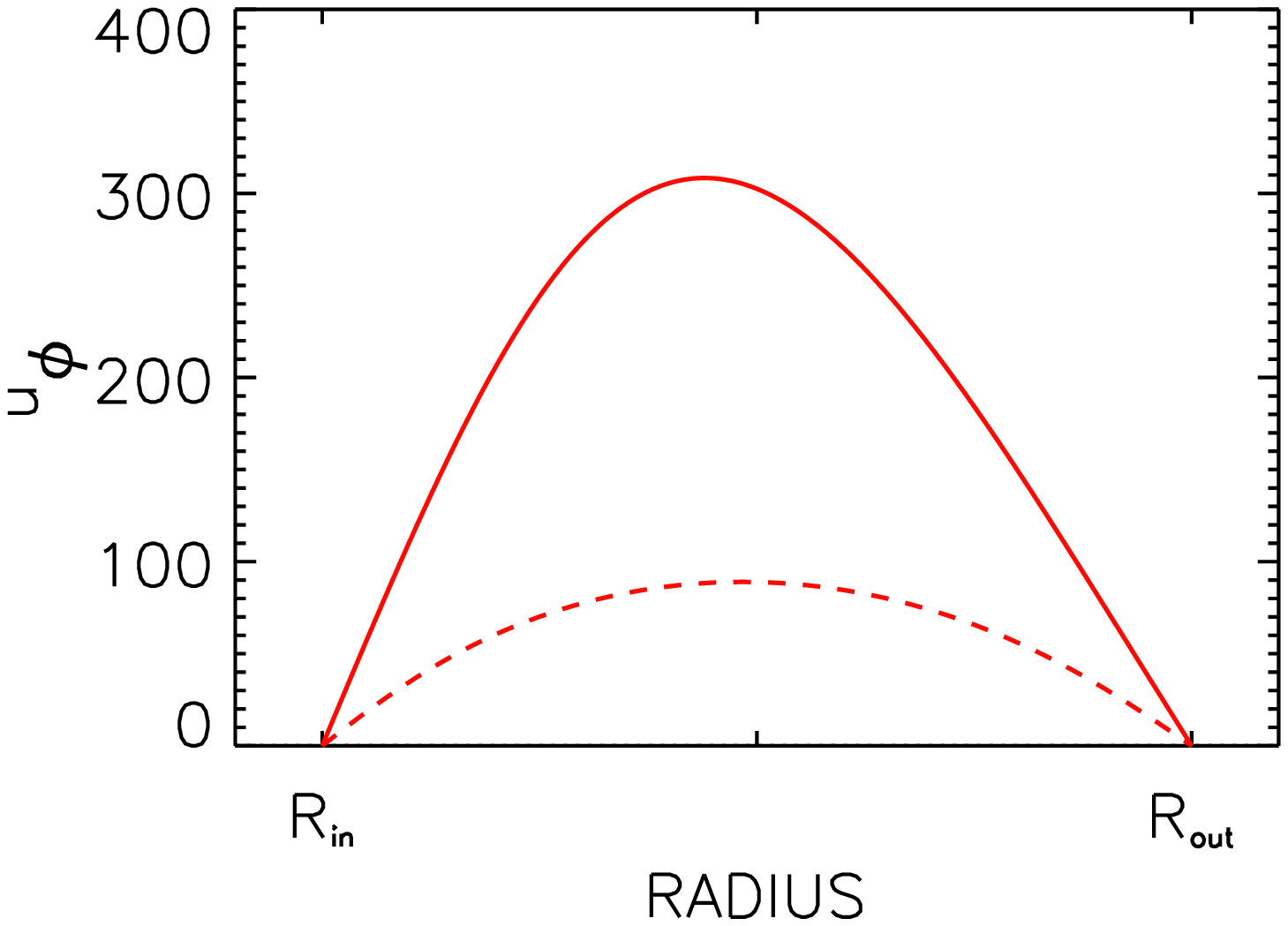}
 \includegraphics[width=0.33\textwidth]{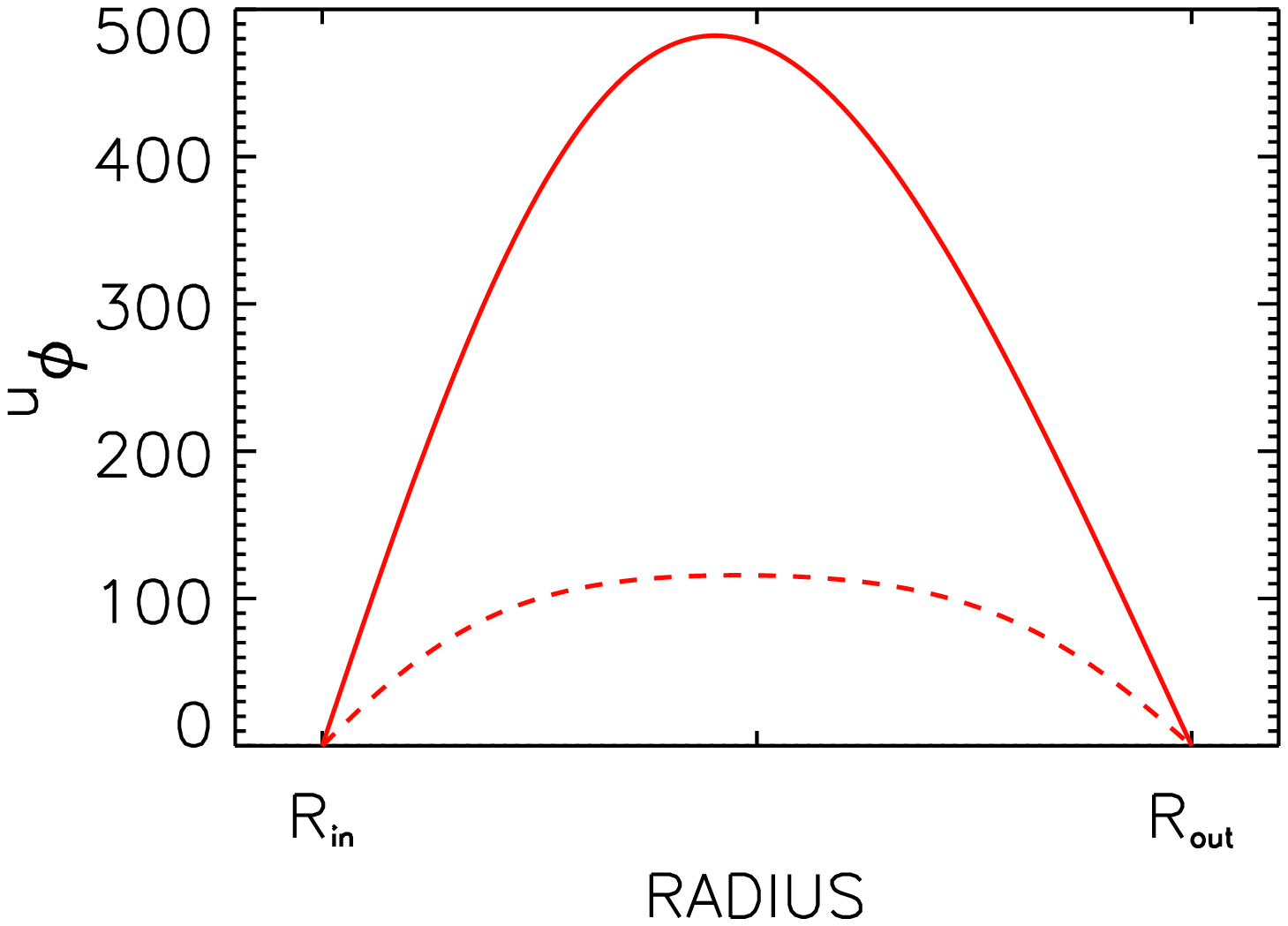}
 \includegraphics[width=0.33\textwidth]{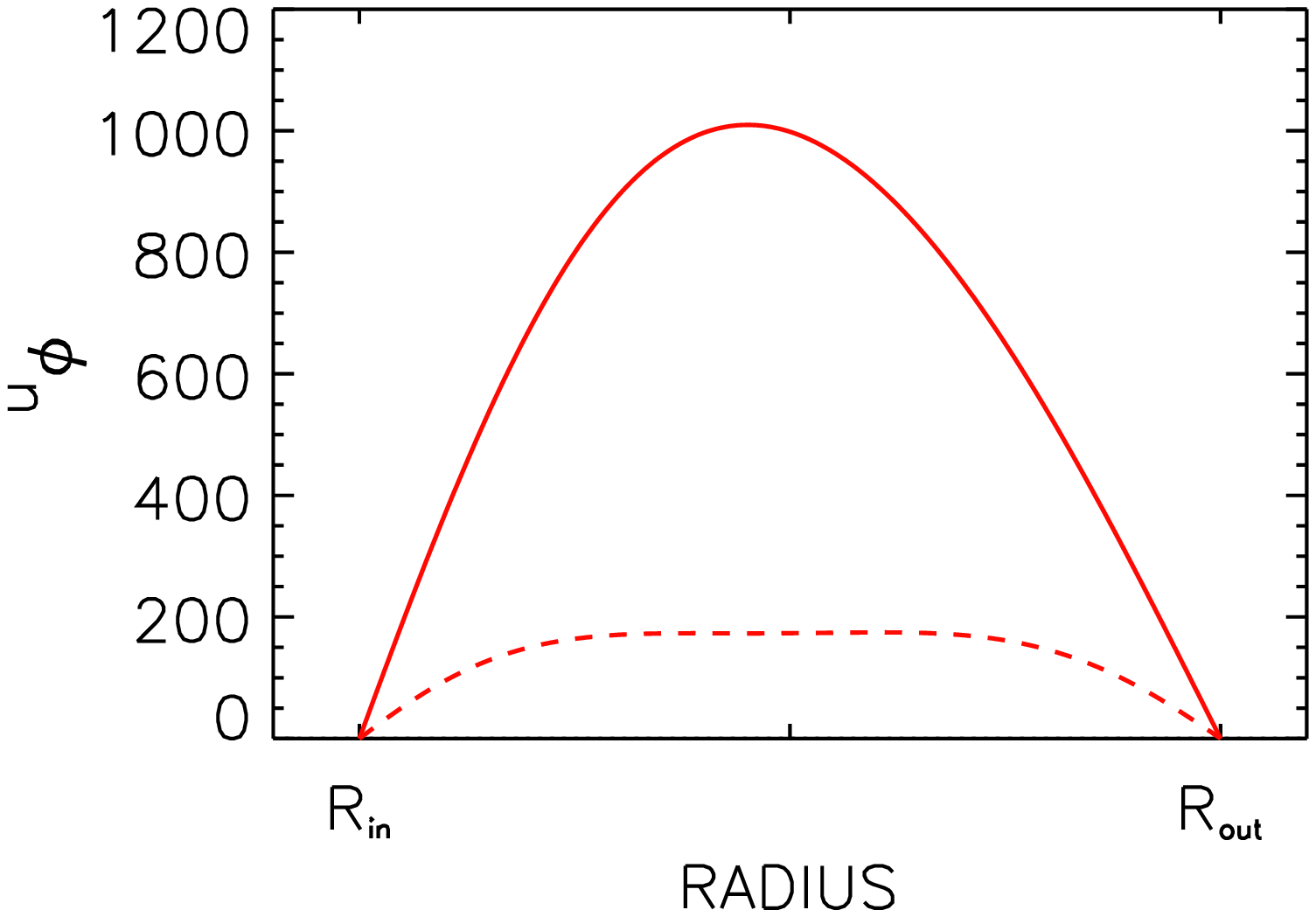}}
}
 \caption{Similar to Fig.~\ref{fig3a} but for $\Pm=0.5$ fixed. From left to right: $\beta=62$, $\beta=32$, $\beta=25$.}
 \label{fig3b} 
\end{figure}

\begin{figure}
 \hbox{
 \includegraphics[width=0.50\textwidth]{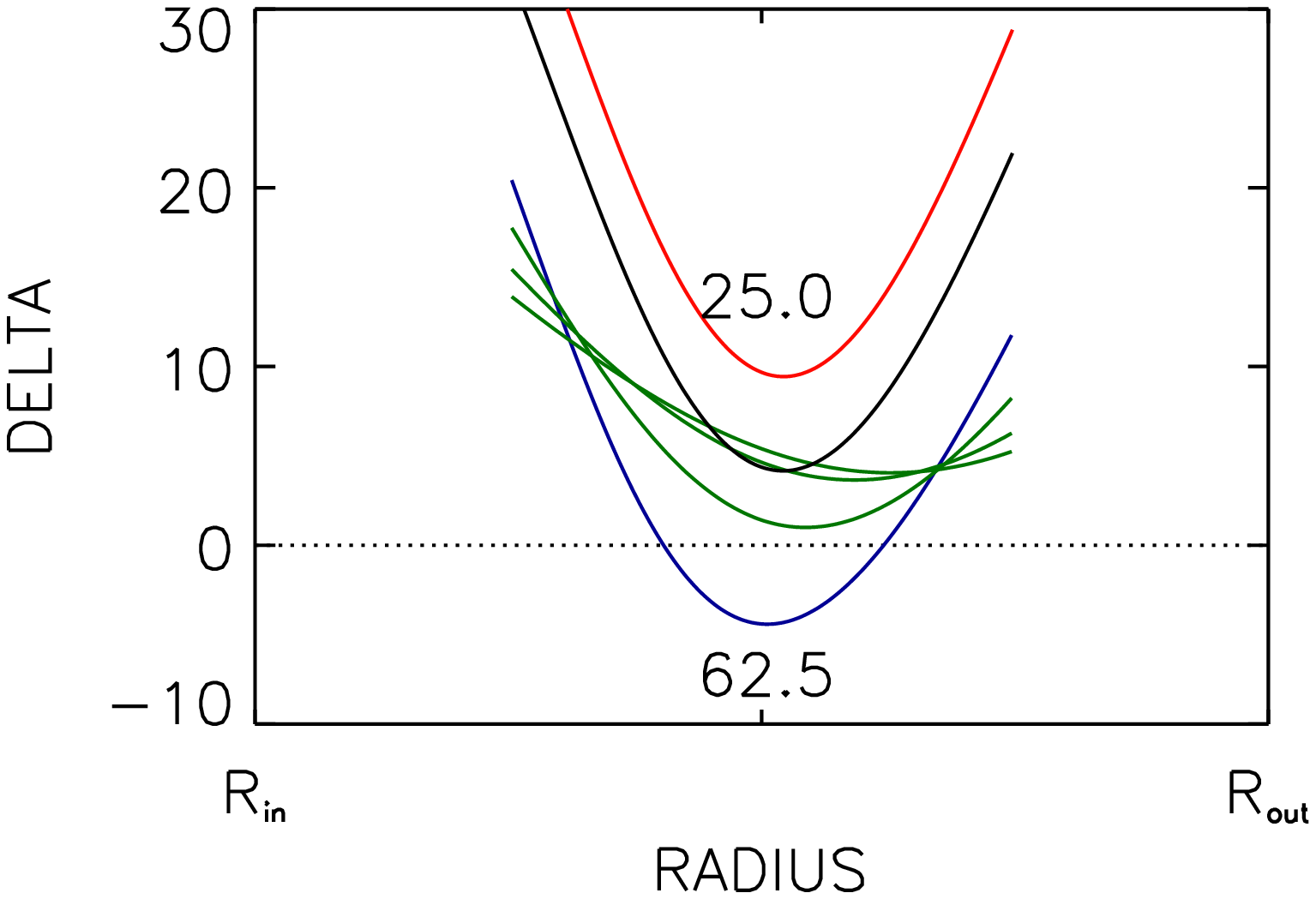}
 \includegraphics[width=0.50\textwidth]{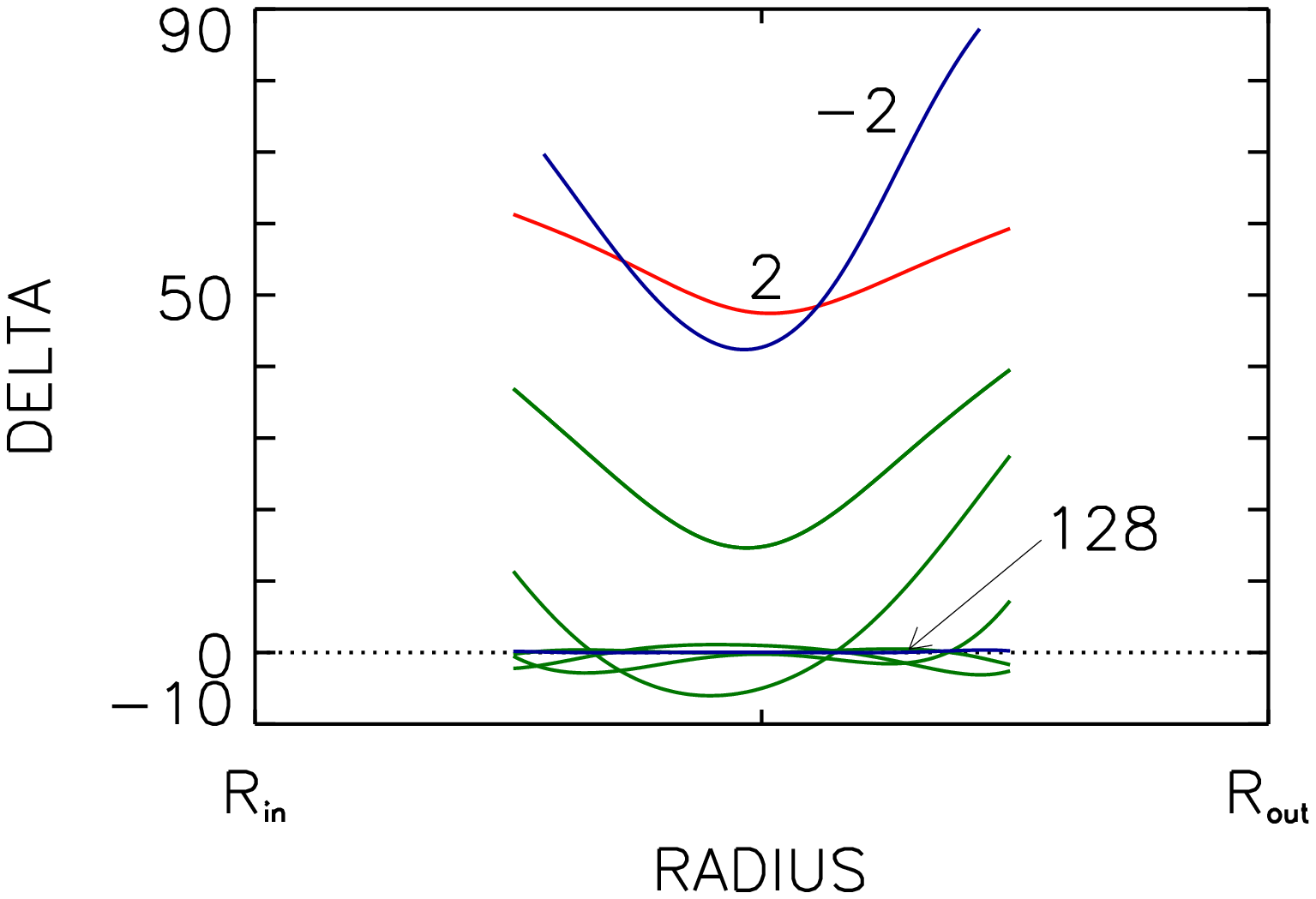}}
 \caption{Phase shifts $\delta$ in degrees according to Eq.~(\ref{delta}) between the azimuthal field and flow components. Left panel: The super-rotating flows of Figs.~\ref{fig3a} and \ref{fig3b}. The blue line corresponds to the model with highest $\beta$ and smallest $\Pm$ (see Fig.~\ref{fig3b}, left). The red line represents the model with $\beta=25$. Right panel: Sub-rotating flows defined in Table \ref{tab2}. The red line denotes $\beta=2$, and the blue line $\beta=-2$.}
 \label{phase}
\end{figure}

Figures \ref{fig3a} and \ref{fig3b} show the radial profiles of $b_\phi$ and $u_\phi$ for the {\em super-rotating} flow with $\mu=128$. Because the solutions contain a free arbitrary factor, only ratios of the components have any physical meaning. The magnetic Prandtl numbers vary between $\Pm=1$ and $\Pm=3$ for fixed $\beta=62$ (Fig.~\ref{fig3a}), and $\beta$ is varied for fixed $\Pm=0.5$ (Fig.~\ref{fig3b}). For all examples one finds $\delta\simeq \pm 10^\circ$, hence the waves of $b_\phi$ and $u_\phi$ are travelling nearly in phase for all $\Pm$ and large $\beta$ (see also \cite{MS19}, their Figs. 8 and 9). For smaller $\beta$ the phase difference grows (left panel of Fig.~\ref{phase}).

Also rotation laws are considered where the outer cylinder rotates slower than the inner one. For a fixed $\Pm=0.5$ some eigenvalues of models with growing magnetic inclination angle $\beta$ are summarized in Table \ref{tab2}.
\begin{table}
\caption{
Eigenvalues for axisymmetric solutions of sub-rotating models with $\mu=\rin=0.5$ and $\Pm=0.1$. $m=0$, minimal Reynolds numbers. 
Models with sub-rotation have always been calculated for perfectly conducting cylinders because of their much easier excitation for small $\Pm$.}
\label{tab2}
\centering
\begin{tabular}{lccccc}
\hline\hline\\

\smallskip
$\beta$ & $\Re_{\rm min}$ &$\Ha$ & $ k R_0$& $\omdr$&$\omega^{\rm R}/\omega_{\rm diff}$\\ \\
\hline
\\
$\pm 2$ & 532 &23.8 &1.65&$\pm 0.069$&$\pm 3.7$\\
4 & 620 & 24.2 &1.43 & 0.095&5.9\\
 8& 990 &23.0& 1.33 &0.068 &6.7\\
 16 & 1898 & 22.1 & 1.35 &0.036&6.9 \\ 
 32 &3883 & 21.9&1.41 & 0.018&7.4\\
64 &8071 &22.4&1.55 &0.0098& 7.9\\
 128&16724 & 23.1 & 1.71& 0.0052&8.7\\\\
 \hline
\end{tabular}
\end{table} 
The radial profiles of $b_\phi$ and $u_\phi$ were calculated for all these flows. The profiles are used for the calculation of the phase differences between the maximum of $b_\phi$ and the maximum of $u_\phi$. The right panel of Fig. \ref{phase} shows the phase shift $\delta$ which also prove to be small for large $\beta$. These waves, therefore, travel in phase along the rotation axis. The result $\delta\simeq 50^\circ $ for small $\beta$ ($|\beta|=2$) demonstrates that for sub-rotating Taylor-Couette flows, i.e. with negative shear, the waves travel in phase, but only for $\beta\gg 1$.

\section{Summary}
The stability problem for axisymmetric and non-axisymmetric perturbations of a magnetized Taylor-Couette flow is analysed where the outer cylinder spins much faster than the inner one (``super-rotation''). The flow is penetrated by a current-free magnetic field of a helical structure with non-vanishing azimuthal and axial components. The ratio $\beta$ of the toroidal and the axial field components plays an important role in determining the stability characteristics of the system. It  is already known that for both extrema $\beta\to 0$ and $\beta\to \infty$ the flow is always stable against axisymmetric perturbations. 

Surprisingly, whether for magnetic Prandtl numbers of order unity the flow is stable or not, basically  depends on the electric boundary conditions. If the cylinders are made with a perfectly conducting material then we did not find a solution for $\Pm\simeq 1$. With insulating boundaries,  however,  solutions exist for all  $\Pm$ (see Fig. \ref{fig2q}). In this case for $\Pm=1$ the flow becomes   unstable for the lowest azimuthal fields if $\beta$  is  large but not too large, i.e.\ $\beta\simeq 60$ (Fig. \ref{figbeta1}, right). Non-axisymmetric modes also exist, but their excitation requires much stronger fields. The domains of instability always possess the characteristic geometry of oblique cones in the ($\Ha/\Re$) plane: for a given supercritical Hartmann number there are a lower and an upper Reynolds number between which the flow is unstable, and similarly for a given supercritical Reynolds number there are a lower and an upper Hartmann number between which the flow is unstable. For axisymmetric patterns the rotation and azimuthal magnetic field must form a magnetic Mach number of order unity; systems with higher magnetic Mach numbers are stable against axisymmetric perturbations, but they may be unstable against non-axisymmetric perturbations. Almost all cosmical objects possess higher Mach numbers, i.e.\ they rotate rapidly compared with their \A-velocity.

The instability pattern always migrates in the axial direction, where the sign of $\beta$ determines the sign of the drift rate. The latter lies between the rotation rate and the diffusion frequency, hence the super-HMRI is basically slower than the standard MRI but faster than any diffusion wave, e.g.\ drifts and waves in dynamo theory.


The sign of the axial drift of the axisymmetric modes depends on the sign of the shear and the sign of the inclination angle $\beta$, hence $\omdr \propto - \beta\ \d\Om/\d R$. For super-rotation, therefore, negative $\beta$ lead to $\omdr<0$, corresponding to a drift anti-parallel to the rotation axis (``equator-ward''). Positive $\beta$ lead to negative drift frequencies $\omdr$ which implies $\dot{z}>0$, i.e.\ pole-ward migration in the northern hemisphere. Equator-ward migration, therefore, complies with negative $\beta$.

For the solar convection zone the quantity $R_0\Om$ is about 800 m/s, so that from Table \ref{tab1} the related phase velocity would become 16 m/s. The axial drift of the super-HMRI therefore exceeds the drift of the solar butterfly diagram ($\simeq$ 1 m/s) by one order of magnitude. This is another formulation of the fact that the time scale of the drift is shorter by a factor of ten than the diffusion time.

These results do not favour an application of the HMRI as a candidate to explain the butterfly phenomenon within the solar activity cycle. The magnetic Prandtl numbers which we used, however, might be too large as they relate to a medium permeated with homogeneous turbulence. For a further test we have checked the phase relation of the azimuthal perturbations of flow and field. It is known from observations of the solar torsional oscillations that they migrate {\em out of phase} toward the equator, i.e.\ the location of $u_\phi=0$ matches the maxima of $b_\phi$. The axial waves of both sorts of HMRI (with sub-rotation and with super-rotation), however, migrate in phase for large $\beta$. Only for small $\beta$ do $u_\phi$ and $b_\phi$ migrate out of phase, in both cases.

As mentioned in the Introduction, due to the positive (negative) latitudinal shear of the solar rotation law at the northern (southern) hemisphere, i.e.  $\cos\theta\ \d\Om/\d \theta>0$, the field geometry parameter $\beta$ should be negative in the northern hemisphere and positive in the southern hemisphere. By the induction of the differential rotation within the solar convection zone we indeed expect $\beta\lsim -10^3$) in the northern hemisphere.

\acknowledgements{D.A. Shalybkov (St. Petersburg) and F. Stefani (Dresden-Rossendorf) are acknowledged for critical readings of the manuscript.}
\appendix
\section{}\label{appA}


Using a short-wave approximation we derived a dispersion relation 
\begin{equation}
\gamma^4+a_2\gamma^2 + \i b_3\gamma +a_4=0
\label{A1}
\end{equation}
(with the denormalized growth rate $\gamma=\i \omega/\Om$) from the linearized equation system (\ref{mhd2}), the solutions of which provide for the stability/instability characteristics of axisymmetric perturbations \citep{RS08}. This result may be used to probe the stability of helical magnetic fields under the presence of differential rotation for ideal flows, i.e.\ $\nu=\eta=0$. The coefficients in (\ref{A1}) were 
\begin{eqnarray}
 a_2= \alpha^2(4-2q) + (4 \alpha^2+\frac{2}{\beta^2}){\widetilde\OmA}^2,\nonumber\\
 b_3= -8 \frac{\alpha^2}{\beta} {\widetilde\OmA}^2,\ \ \ \ \ \ \ \ \ a_4= \frac{{\widetilde\OmA}^4}{\beta^4} -2 q \frac{\alpha^2}{\beta^2} {\widetilde\OmA}^2.
 \label{coeff} 
\end{eqnarray}
Here $\alpha=k_z/|\vec{k}|$ is the axial wave number (normalized with the total wave number $k$). The shear is defined by the radial rotation law $\Om\propto R^{-q}$. For $q=2$ (potential flow) and $q=1$ (quasi-uniform flow) rotation profiles with negative shear are defined while negative $q$ describe super-rotation. 
 We shall numerically determine the scaling of the growth rate as the real part of the complex $\gamma$ with the magnetic system parameter $\widetilde\OmA=\OmAf/\Om$ which is the inverse of the azimuthal magnetic Mach number $\Mm$. $\OmAf=B_\phi/\sqrt{\mu_0\rho}R$ and $\OmAz=k_z B_z/\sqrt{\mu_0\rho}$ are the \A~frequencies of the azimuthal and the axial magnetic field components. Following \cite{KSF12} the ratio $\beta$ of the azimuthal field and the axial field may be written as
 $\beta= \OmAf/\OmAz$ (corresponding to but not identical with (\ref{beta})). 

\begin{figure}
 \includegraphics[width=0.52\textwidth]{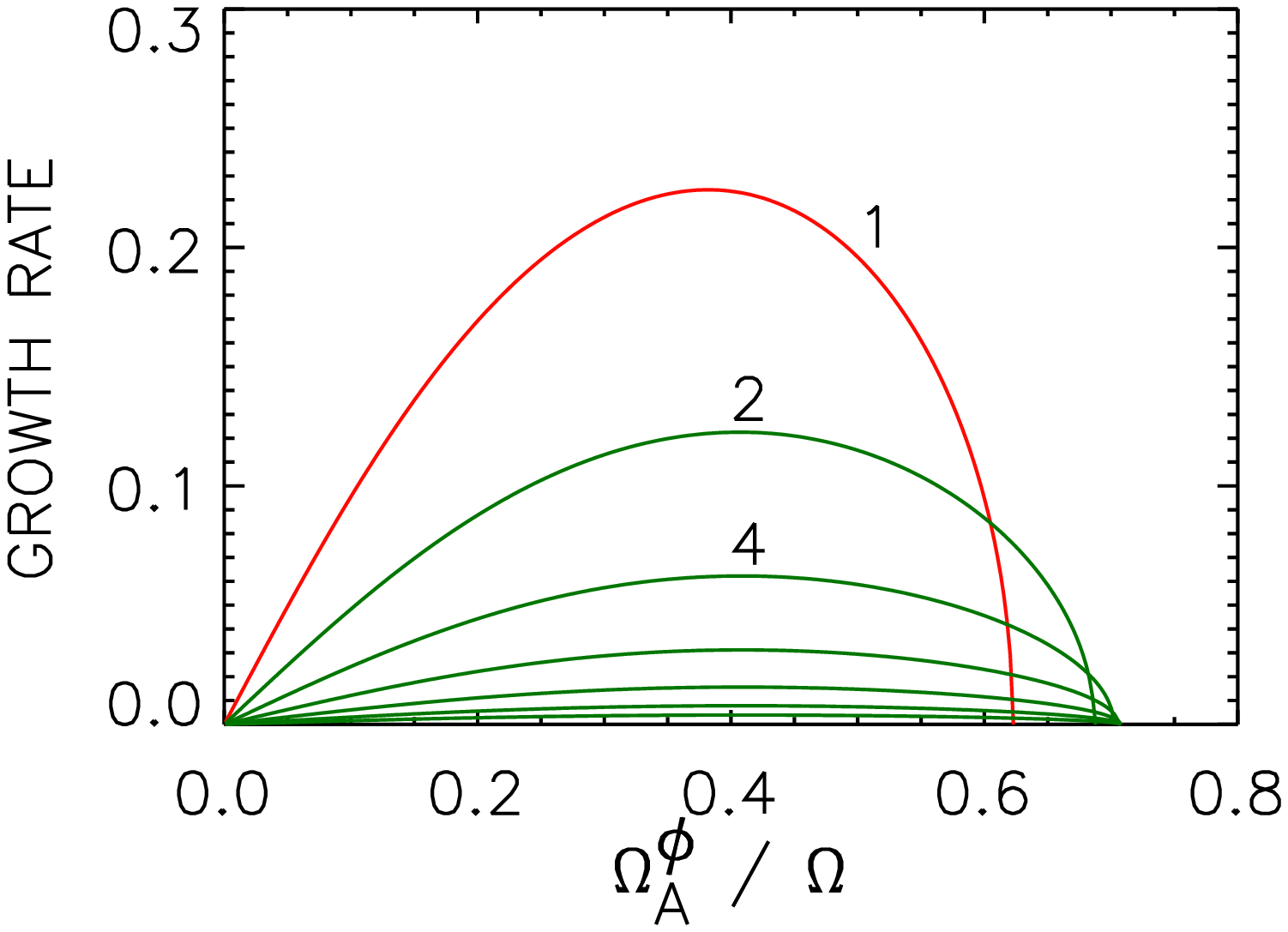}
 \includegraphics[width=0.55\textwidth]{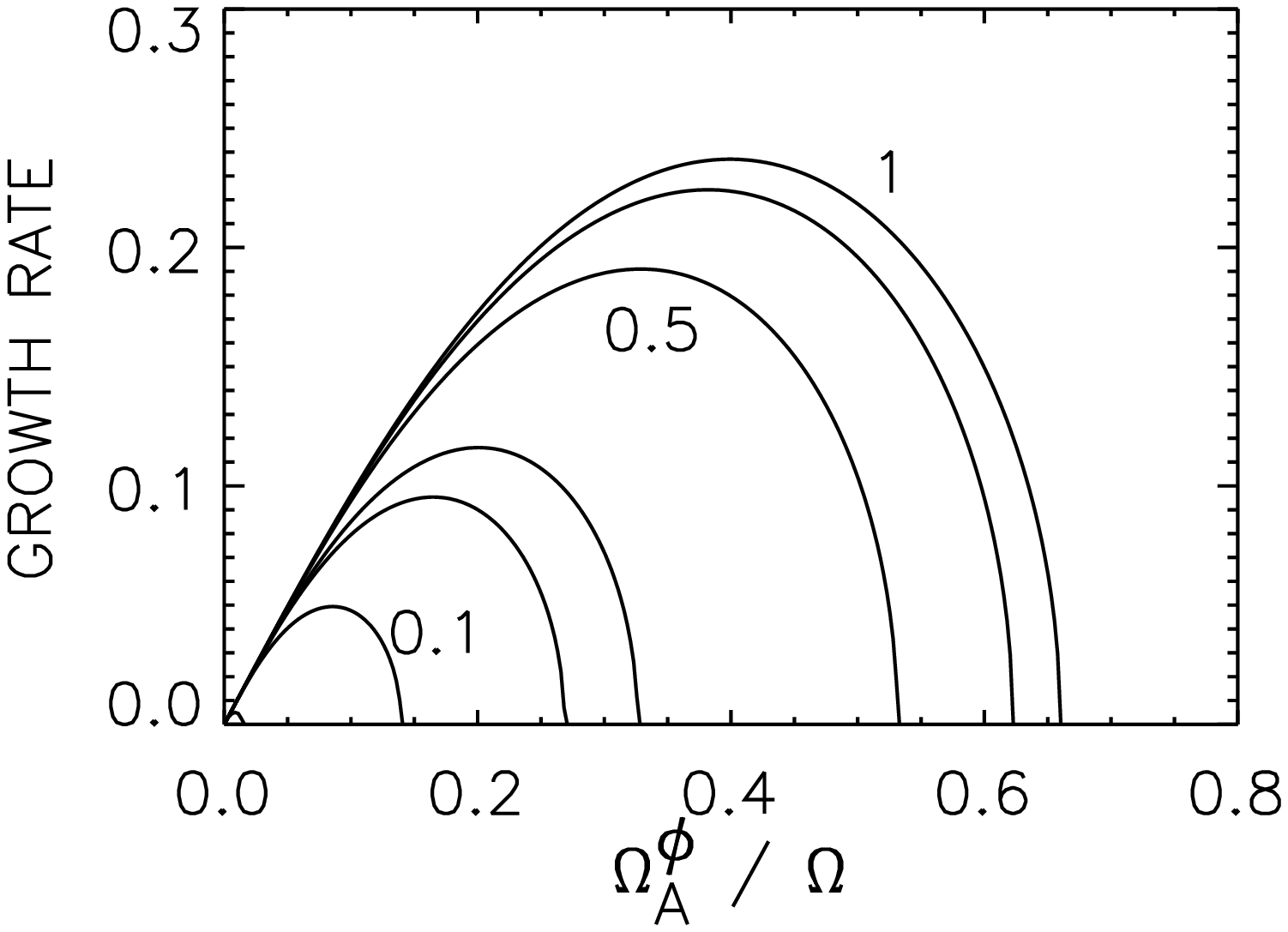}
 \caption{Normalized growth rates $\gamma$ of  classical  HMRI with $m=0$. The horizontal coordinate is $\widetilde\OmA$. Left: Variation of $\beta$ (marked), $\alpha=0.75$. Right: Variation of $\alpha$ (marked) for $\beta=1$. 
 The rotation profile  is $\Om\propto R^{-1}$, i.e. $q=1$.} 
 \label{ideal2} 
\end{figure}

Here we shall only work for $q=\pm 1$. The ratio $\alpha$ must be considered as a free parameter which varies as $0<\alpha <1$. Figure \ref{ideal2} demonstrates the existence of positive growth rates for the rotation law $\Om\propto R^{-1}$ (i.e.\ for uniform  rotation velocity) up to a certain upper limit of $\OmAf/\Om$ corresponding to a magnetic Mach number of $\Mm\simeq 1.4$. The flow becomes unstable if its Mach number exceeds this value -- or in other words, if its rotation is rapid enough. The instability condition for ideal flows obviously represents only the {\em lower} branch of the complete line of marginal stability of real fluids. The numerical value  approaches the Mach number for $\Pm=1$ of the { lower} branch of the instability cone for quasi-uniform linear rotation. Obviously, the upper branch of the instability cone which stabilizes the flow for higher Reynolds numbers is basically due to finite diffusivities. We note that within the short-wave approximation the upper branch with the maximally possible Reynolds number will basically not be provided.

We find that the axisymmetric HMRI with negative shear also exists for ideal fluids. This, however, is not true for positive shear. The relation (\ref{A1}) for $q<0$ does not provide solutions with positive real part of $\gamma$. The axisymmetric super-HMRI, therefore, is a diffusion-originated instability which only exists for non-zero diffusivities $\nu$ and $\eta$. Both branches of its instability cone are thus due to diffusion processes.

The left panel of Fig. \ref{ideal2} demonstrates that the dimensional maximal growth rate 
\begin{equation}
\gamma\propto \frac{\OmAf}{\beta} = \OmAz
\label{growth}
\end{equation}
 is only determined by the { axial} magnetic field. On the other hand, the system is unstable for all $\OmAf/\Om\lsim 0.7$, i.e.\ for all magnetic Mach numbers exceeding 1.4. One also finds a very weak influence of $\beta$ on the critical ratio $\OmAf/\Om$ for marginal excitation ($\gamma=0$). The lower branches of the instability cones of real flows should thus be almost identical for all $\beta$ (see Fig. \ref{figbeta1}). 
 
 The right panel of Fig. \ref{ideal2} demonstrates a rather strong monotonous influence of the normalized axial wave number onto the growth rate profiles. As also the horizontal coordinate contains $k_z$, the plot reflects mainly the influence of the radial wave number, i.e.\ the width of the cylinder gap.

\bibliographystyle{jpp}
\bibliography{superamri.bib}

\begin{thebibliography}{22}
\expandafter\ifx\csname natexlab\endcsname\relax\def\natexlab#1{#1}\fi
\def\au#1{#1} \def\ed#1{#1} \def\yr#1{#1}\def\at#1{#1}\def\jt#1{\textit{#1}}
  \def\bt#1{#1}\def\bvol#1{\textbf{#1}} \def\vol#1{#1} \def\pg#1{#1}
  \def\publ#1{#1}\def\arxiv#1{#1}\def\org#1{#1}\def\st#1{\textit{#1}}

\bibitem[{Acheson}(1978)]{A78}
{\sc \au{{Acheson}, D.~J.}} \yr{1978}  \at{{On the instability of toroidal
  magnetic fields and differential rotation in stars}}.  \jt{Philosophical
  Transactions of the Royal Society of London Series A}  \bvol{289},
  \pg{459--500}.

\bibitem[{Caspary} {\em et~al.\/}(2018){Caspary}, {Choi}, {Ebrahimi}, {Gilson},
  {Goodman} \& {Ji}]{CC18}
{\sc \au{{Caspary}, K.~J.}, \au{{Choi}, D.}, \au{{Ebrahimi}, F.}, \au{{Gilson},
  E.~P.}, \au{{Goodman}, J.} \& \au{{Ji}, H.}} \yr{2018}  \at{{Effects of axial
  boundary conductivity on a free Stewartson-Shercliff layer}}.  \jt{\pre}
  \bvol{97}~(6),  \pg{063110}.

\bibitem[{Choi} {\em et~al.\/}(2019){Choi}, {Ebrahimi}, {Caspary}, {Gilson},
  {Goodman} \& {Ji}]{CE19}
{\sc \au{{Choi}, D.}, \au{{Ebrahimi}, F.}, \au{{Caspary}, K.~J.}, \au{{Gilson},
  E.~P.}, \au{{Goodman}, J.} \& \au{{Ji}, H.}} \yr{2019}  \at{{Nonaxisymmetric
  simulations of the Princeton magnetorotational instability experiment with
  insulating and conducting axial boundaries}}.  \jt{\pre}  \bvol{100}~(3),
  \pg{033116}.

\bibitem[{Gellert} {\em et~al.\/}(2012){Gellert}, {R{\"u}diger} \&
  {Schultz}]{GR12}
{\sc \au{{Gellert}, M.}, \au{{R{\"u}diger}, G.} \& \au{{Schultz}, M.}}
  \yr{2012}  \at{{The angular momentum transport by standard MRI in
  quasi-Kepler cylindrical Taylor-Couette flows}}.  \jt{\aa}  \bvol{541},
  \pg{A124}.

\bibitem[{Hollerbach} \& {R{\"u}diger}(2005)]{HR05}
{\sc \au{{Hollerbach}, R.} \& \au{{R{\"u}diger}, G.}} \yr{2005}  \at{{New Type
  of Magnetorotational Instability in Cylindrical Taylor-Couette Flow}}.
  \jt{Physical Review Letters}  \bvol{95}~(12),  \pg{124501}.

\bibitem[{Ji} {\em et~al.\/}(2001){Ji}, {Goodman} \& {Kageyama}]{JG01}
{\sc \au{{Ji}, H.}, \au{{Goodman}, J.} \& \au{{Kageyama}, A.}} \yr{2001}
  \at{{Magnetorotational instability in a rotating liquid metal annulus}}.
  \jt{\mnras}  \bvol{325},  \pg{L1--L5}.

\bibitem[{Kirillov}(2013)]{K13}
{\sc \au{{Kirillov}, O.}} \yr{2013} {\em {Nonconservative Stability Problems of
  Modern Physics}\/}.  \publ{{De Gruyter, Berlin}}.

\bibitem[{Kirillov}(2017)]{K17}
{\sc \au{{Kirillov}, O.~N.}} \yr{2017}  \at{{Singular diffusionless limits of
  double-diffusive instabilities in magnetohydrodynamics}}.  \jt{Proceedings of
  the Royal Society of London Series A}  \bvol{473}~(2205),  \pg{20170344},
  \arxiv{arXiv: 1610.06970}.

\bibitem[{Kirillov} {\em et~al.\/}(2012){Kirillov}, {Stefani} \&
  {Fukumoto}]{KS12}
{\sc \au{{Kirillov}, O.~N.}, \au{{Stefani}, F.} \& \au{{Fukumoto}, Y.}}
  \yr{2012}  \at{{A Unifying Picture of Helical and Azimuthal Magnetorotational
  Instability, and the Universal Significance of the Liu Limit}}.  \jt{\apj}
  \bvol{756},  \pg{83}.

\bibitem[{Kirillov} {\em et~al.\/}(2014){Kirillov}, {Stefani} \&
  {Fukumoto}]{KSF12}
{\sc \au{{Kirillov}, O.~N.}, \au{{Stefani}, F.} \& \au{{Fukumoto}, Y.}}
  \yr{2014}  \at{{Local instabilities in magnetized rotational flows: a
  short-wavelength approach}}.  \jt{Journal of Fluid Mechanics}  \bvol{760},
  \pg{591--633}.

\bibitem[{Komm} {\em et~al.\/}(2016){Komm}, {Howe} \& {Hill}]{KHH16}
{\sc \au{{Komm}, R.}, \au{{Howe}, R.} \& \au{{Hill}, F.}} \yr{2016} {Subsurface
  Zonal and Meridional Flows from SDO/HMI}.  \bt{In {\em SDO 2016: Unraveling
  the Sun's Complexity\/} (ed. \ed{W.~Dean {Pesnell} \& Barbara {Thompson}})},
  \pg{p.~55}.

\bibitem[{Liu} {\em et~al.\/}(2006){Liu}, {Goodman}, {Herron} \& {Ji}]{LG06}
{\sc \au{{Liu}, W.}, \au{{Goodman}, J.}, \au{{Herron}, I.} \& \au{{Ji}, H.}}
  \yr{2006}  \at{{Helical magnetorotational instability in magnetized
  Taylor-Couette flow}}.  \jt{\pre}  \bvol{74}~(5),  \pg{056302}.

\bibitem[{Mamatsashvili} {\em et~al.\/}(2019){Mamatsashvili}, {Stefani},
  {Hollerbach} \& {R{\"u}diger}]{MS19}
{\sc \au{{Mamatsashvili}, G.}, \au{{Stefani}, F.}, \au{{Hollerbach}, R.} \&
  \au{{R{\"u}diger}, G.}} \yr{2019}  \at{{Two types of axisymmetric helical
  magnetorotational instability in rotating flows with positive shear}}.
  \jt{Physical Review Fluids}  \bvol{4}~(10),  \pg{103905}.

\bibitem[{R{\"u}diger} {\em et~al.\/}(2018{\natexlab{{\em a\/}}}){R{\"u}diger},
  {Gellert}, {Hollerbach}, {Schultz} \& {Stefani}]{RGH18}
{\sc \au{{R{\"u}diger}, G.}, \au{{Gellert}, M.}, \au{{Hollerbach}, R.},
  \au{{Schultz}, M.} \& \au{{Stefani}, F.}} \yr{2018{\natexlab{{\em a\/}}}}
  \at{{Stability and instability of hydromagnetic Taylor-Couette flows}}.
  \jt{\physrep}  \bvol{741},  \pg{1--89},  \arxiv{arXiv: 1703.09919}.

\bibitem[{R{\"u}diger} \& {Schultz}(2008)]{RS08}
{\sc \au{{R{\"u}diger}, G.} \& \au{{Schultz}, M.}} \yr{2008}  \at{{Helical
  magnetorotational instability of Taylor-Couette flows in the Rayleigh limit
  and for quasi-Kepler rotation}}.  \jt{Astronomische Nachrichten}  \bvol{329},
   \pg{659}.

\bibitem[{R{\"u}diger} {\em et~al.\/}(2018{\natexlab{{\em b\/}}}){R{\"u}diger},
  {Schultz}, {Gellert} \& {Stefani}]{RSG18}
{\sc \au{{R{\"u}diger}, G.}, \au{{Schultz}, M.}, \au{{Gellert}, M.} \&
  \au{{Stefani}, F.}} \yr{2018{\natexlab{{\em b\/}}}}  \at{{Azimuthal
  magnetorotational instability with super-rotation}}.  \jt{Journal of Plasma
  Physics}  \bvol{84}~(1),  \pg{735840101}.

\bibitem[{R{\"u}diger} {\em et~al.\/}(2018{\natexlab{{\em c\/}}}){R{\"u}diger},
  {Schultz}, {Stefani} \& {Hollerbach}]{RSS18}
{\sc \au{{R{\"u}diger}, G.}, \au{{Schultz}, M.}, \au{{Stefani}, F.} \&
  \au{{Hollerbach}, R.}} \yr{2018{\natexlab{{\em c\/}}}}
  \at{{Magnetorotational instability in Taylor-Couette flows between cylinders
  with finite electrical conductivity}}.  \jt{Geophysical and Astrophysical
  Fluid Dynamics}  \bvol{112},  \pg{301--320},  \arxiv{arXiv: 1804.01501}.

\bibitem[{R{\"u}diger} \& {Zhang}(2001)]{RZ01}
{\sc \au{{R{\"u}diger}, G.} \& \au{{Zhang}, Y.}} \yr{2001}  \at{{MHD
  instability in differentially-rotating cylindric flows}}.  \jt{\aa}
  \bvol{378},  \pg{302--308}.

\bibitem[{Seilmayer} {\em et~al.\/}(2012){Seilmayer}, {Stefani}, {Gundrum},
  {Weier}, {Gerbeth}, {Gellert} \& {R{\"u}diger}]{SS12}
{\sc \au{{Seilmayer}, M.}, \au{{Stefani}, F.}, \au{{Gundrum}, T.}, \au{{Weier},
  T.}, \au{{Gerbeth}, G.}, \au{{Gellert}, M.} \& \au{{R{\"u}diger}, G.}}
  \yr{2012}  \at{{Experimental Evidence for a Transient Tayler Instability in a
  Cylindrical Liquid-Metal Column}}.  \jt{Physical Review Letters}
  \bvol{108}~(24),  \pg{244501}.

\bibitem[{Shalybkov} {\em et~al.\/}(2002){Shalybkov}, {R{\"u}diger} \&
  {Schultz}]{SR02}
{\sc \au{{Shalybkov}, D.~A.}, \au{{R{\"u}diger}, G.} \& \au{{Schultz}, M.}}
  \yr{2002}  \at{{Nonaxisymmetric patterns in the linear theory of MHD
  Taylor-Couette instability}}.  \jt{\aap}  \bvol{395},  \pg{339--343},
  \arxiv{arXiv: astro-ph/0207331}.

\bibitem[{Stefani} {\em et~al.\/}(2006){Stefani}, {Gundrum}, {Gerbeth},
  {R{\"u}diger}, {Schultz}, {Szklarski} \& {Hollerbach}]{SG06}
{\sc \au{{Stefani}, F.}, \au{{Gundrum}, T.}, \au{{Gerbeth}, G.},
  \au{{R{\"u}diger}, G.}, \au{{Schultz}, M.}, \au{{Szklarski}, J.} \&
  \au{{Hollerbach}, R.}} \yr{2006}  \at{{Experimental Evidence for
  Magnetorotational Instability in a Taylor-Couette Flow under the Influence of
  a Helical Magnetic Field}}.  \jt{Physical Review Letters}  \bvol{97}~(18),
  \pg{184502}.

\bibitem[{Velikhov}(1959)]{V59}
{\sc \au{{Velikhov}, E.}} \yr{1959}  \at{{Stability of an ideally conducting
  liquid flowing between cylinders rotating in a magnetic field}}.  \jt{Soviet.
  Phys. JETP}  \bvol{36},  \pg{1389--1404}.

\end{thebibliography}

\end{document}